\documentclass{elsarticle}
\usepackage{amssymb,latexsym,graphicx,hyperref,color}

\usepackage{diagrams} 
 
\newtheorem{theorem}{Theorem}
\newtheorem{proposition}[theorem]{Proposition}
\newtheorem{lemma}[theorem]{Lemma}

\newcommand{\id}{id}
\newcommand{\Pos}{\mathcal{P}}
\newcommand{\Neg}{\mathcal{N}}
\newcommand{\bigcomma}{  \mathbin{\:\raisebox{.5ex}{\LARGE\bf ,}\:} } 
\newcommand{\slot}{\mbox{-}} 
\newcommand{\nil}{\mathbf{nil}}
\newcommand{\Sep}{\mathrel{\:*\:}}
\newcommand{\SEP}{\mathrel{\;*\;}}
\newcommand{\cell}[3]{#1\stackrel{.}{\mapsto} #2,#3} 
\newcommand{\bdot}{\,.\,}  
\newcommand{ \rt }{\mathrel{\sqsubseteq}} 
\newcommand{ \rf }{\mathrel{\sqsupseteq}} 
\newcommand{\typ}{\sigma} 
\newcommand{ \ty }{\!:\!}  
\newcommand{\LL}{\langle} 
\newcommand{\RR}{\rangle}
\newcommand{\Lam}{ \mathord{{\Lambda}} } 
\newcommand{\gr}{\mathsf{gr}}
\newcommand{\rg}{\mathsf{rg}}
\newcommand{\univ}{\mathsf{unim}}
\newcommand{\exi}{\mathsf{exim}}
\newcommand{\simi}{\mathsf{sim}} 
\newcommand{\sims}{\mathsf{s}} 
\newcommand{\psim}{\mathsf{psim}} 
\newcommand{\comp}{\mathsf{comp}}
\newcommand{\rev}{\mathsf{rev}} 

\newcommand{\homm}{\mathbin{\Rightarrow}} 
\newcommand{\ap}{\mathsf{ap_f}} 
\newcommand{\cur}{\mathsf{cur_f}}

\newcommand{\homs}{\mathbin{\ltimes}} 
\newcommand{\aps}{\mathsf{ap_x}} 
\newcommand{\curs}{\mathsf{cur_x}}
\newcommand{\uncurs}{\mathsf{uncur_x}}

\newcommand{\homi}{\mathbin{\leadsto}} 
\newcommand{\api}{\mathsf{ap_i}} 
\newcommand{\curi}{\mathsf{cur_i}}
\newcommand{\uncuri}{\mathsf{uncur_i}}

\newcommand{\homt}{\mathbin{\looparrowright}} 
\newcommand{\apt}{\mathsf{ap_t}} 
\newcommand{\curt}{\mathsf{cur_t}}

\newcommand{\plambda}{\mathord{\raisebox{.45ex}{$\chi$}}} 
\newcommand{\patarrow}{\mathrel{\mapsto}} 
\newcommand{\typarrow}{\rightarrow} 
\newcommand{\arrow}{\mathbin{\longrightarrow}} 
\newcommand{\mor}[3]{ #1 : #2 \arrow #3} 

\newcommand{\BigPatarrow}{\mbox{\Large$\mathbf{\patarrow}$}} 

\newcommand{\E}{ \mathord{{\mathbb E}}}
\newcommand{\A}{ \mathord{{\mathbb A}}}
\newcommand{\Ni}{\mathord{\ni}} 

\newcommand{\F}{{\mathcal F }} 
\newcommand{\R}{{\mathcal R }} 
\newcommand{\PT}{{\mathcal T }} 
\newcommand{\Fi}{} 
\newcommand{\Ri}{} 
\newcommand{\PTi}{} 

\newcommand{ \G }{\Gamma}
\newcommand{ \tg }{{ \mathrel{\;\vdash}\;}} 
\newcommand{ \se }{ { [ \! [ } }              
\newcommand{ \es }{ { ] \! ] } }              

\newcommand{\one}{\mathbf{1}}
\newcommand{ \caten }{\mathbin{\mbox{\small $+\!\!+$}}}
\newcommand{\Id}{\mathsf{id}} 

\newcommand{\co}{\mathbin{\mbox{\small\upshape\textbf{;}}}} 

\newcommand{\Sc}{\mathbin{{\stackrel{\mbox{\tiny$\circ$}}{,}}}}
\newcommand{\op}[1]{\widetilde{#1}} 
\newcommand{\postop}[1]{(#1)^{\sim}} 
\newcommand{\Gr}{\mathord{\sf Gr}} 
\newcommand{\Rg}{\mathord{\sf Rg}} 
\renewcommand{\ss}{\mathrel{\subseteq}}
\newcommand{\sps}{\mathrel{\supseteq}}
\newcommand{\ssps}{\mathrel{\,\supseteq\,}} 
\newcommand{\rsps}{\vcenter{\hbox{\tiny $\supseteq $}}}
\newcommand{\rpol}{\vcenter{\hbox{\tiny $\preceq $}}} 
\newcommand{\cat}[1]{\mbox{\upshape\textbf{#1}}} 

\newcommand{\Mofun}{\cat{Poset}} 
\newcommand{\Idl}{\cat{Idl}} 
\newcommand{\Tran}{\cat{Tran}}    
\newcommand{\Fun}{\cat{Fun}}
\newcommand{\Rel}{\cat{Rel}}
\newcommand{\U}{{\mathcal U}} 
\newcommand{\Uop}{\op{\mathcal U}} 
\newcommand{\dt}[1]{{\em #1}} 
\newcommand{ \pol }{\preceq} 
\newcommand{ \pog }{\succeq} 
\newcommand{ \all }{\forall\,}
\newcommand{ \some }{\exists\,}
\newcommand{ \equ }{\mathrel{  \equiv  }}

\newenvironment{calcu}{                   
  \begin{center}$ \begin{array}{rll}
  }{
  \end{array} $ \end{center}    }
\newcommand{ \fln }[1]{ & #1             \protect\\*[.5ex] }
\newcommand{ \pln }[3]{%
   #1  &\quad\; \mbox{{#3}}          \protect\\*[.5ex]
       & \multicolumn{2}{l}{#2}        \protect\\*[.5ex] }

\newenvironment{calcux}{ 
  \begin{center}$ \begin{array}{rll}
  }{
  \end{array} $ \end{center}  }
 \newcommand{ \flnx }[1]{   & #1 &        \protect\\*[-.5ex] }
\newcommand{ \plnx }[3]{%
    #1 &   & \;      \mbox{#3} \protect\\*[-.8ex]
       & #2 &                \protect\\*[-.8ex]  }

\begin{document}
\title{Towards Patterns for Heaps and Imperative Lambdas\\[.5ex]
{\small to appear in Elsevier JLAMP, DOI: 10.1016/j.jlamp.2015.10.008\\[-.5ex]
\mbox{\copyright 2015 This manuscript version is made available under the CC-BY-NC-ND 4.0 license.}
}}

\author{David A. Naumann} 
\address{Stevens Institute of Technology,  USA
\vspace*{-6ex} 
}

\begin{abstract}
In functional programming, point-free relation calculi have been fruitful for general theories of program construction, but for specific applications pointwise expressions can be more convenient and comprehensible. In imperative programming, refinement calculi have been tied to pointwise expression in terms of state variables, with the curious exception of the ubiquitous but invisible heap. To integrate pointwise with point-free, de Moor and Gibbons \cite{deMoor:Gib} extended lambda calculus with non-injective pattern matching interpreted using relations.  This article gives a semantics of that language using ``ideal relations'' between partial orders, and a second semantics using predicate transformers. The second semantics is motivated by its potential use with separation algebra, for pattern matching in programs acting on the heap. Laws including lax beta and eta are proved in these models and a number of open problems are posed.  
\end{abstract}

\maketitle

Dedicated to Jos\'{e} Nuno Oliveira on the occasion of his 60th birthday.

\section{Introduction}\label{sect:intro}

An important idea in the mathematics of program construction is to embed the programming language of interest into a richer language with additional features that are useful for writing specifications and for reasoning.  
Functional programs can be embedded in the calculus of relations, which 
provides two key benefits: converse functions as specifications and intersection of specifications.
An example of the first benefit is parsing.  Let $\mor{show}{Tree}{String}$ be the function that maps an ordered tree with strings at its leaves to the ``inorder'' catenation of the leaves.  Its converse, $show^o$, is a relation but not a function.  
One  seeks to derive, by algebraic reasoning in the calculus of relations, a total function $\mathit{parse}$ such that $show^o \sps \mathit{parse}$.
See Bird and de Moor~\cite{BirdDeM} for many more examples.
Imperative programs can be embedded in a refinement calculus~\cite{Back:book,Morgan:book}, by augmenting the language with assumptions and angelic choice, or ``specification statements'' in some other form. These can be  modeled using weakest precondition predicate transformers.
An imperative program $prog$ satisfies specification $spec$ just if 
$spec \rt prog$ where $\rt$ is the pointwise order on predicate transformers,
and again one seeks to derive $prog$ from $spec$.

Many authors have pointed out useful and elegant aspects of the calculus of relations for programming.
Relations cater for the development of general theory by facilitating a ``point free'' style in which algebraic 
calculation is not encumbered by manipulation of bound variables and substitutions
(e.g., see \cite{Oliveira08ESC}).

Although pointfree style is elegant and effective for development of general theory, it can be awkward and cryptic for developing and expressing specific algorthms.  Functional programmers tend to prefer a mix of pointfree and pointwise expressions, ``pointwise'' meaning the use of variables and other expressions that denote data elements ---application rather than composition.
Pointwise reasoning involves logical quantifiers and is the norm in imperative program construction.
For example, refinement laws for assignment statements involve conditions on free variables, and specifications are expressed in terms of state variables and formulas with quantifiers.

Conventional pattern matching can help raise the abstraction level in pointwise programs, by directly expressing data structure of interest. 
Non-injective patterns have been proposed by de Moor and Gibbons \cite{deMoor:Gib} as a way to achieve pointwise programming with relations.
Imperative programmers draw graphs to express patterns of pointer structure, but their programs are written in impoverished notation that amounts to little more than load and store instructions.

This article contributes to the long term goal of a unified theory of programming in which one may move freely between pointwise or pointfree reasoning as suits the occasion.
For example, requirements might be formalized in a transparent pointwise specification that is then transformed to a pointfree equivalent from which an efficient solution is derived by algebraic calculation.
A unified theory will also enable effective mixes of functional, imperative, and other styles both in program structure and in reasoning.

This article describes one approach to a programming calculus integrating functional and imperative styles, addressing some aspects of pointwise and pointfree reasoning.
Some of the technical results were published in a conference paper by the author \cite{Naumann01b},
from which much of the material is adapted.  
The introductory sections have been rewritten using different examples. 
This article provides full details of the main semantic definitions and some results only mentioned sketchily in the conference paper, namely beta and eta laws.
We cannot expect beta and eta equalities to hold unrestrictedly, as they fail already in
by-value functional languages. Inequational laws are mentioned but not
proved in \cite{deMoor:Gib} and \cite{Naumann01b}.
Here we prove weak beta and eta laws for both relational and predicate transformer
semantics.  
We also pose several open problems.

\paragraph{Outline}

The rest of this article is organized as follows.
Section~\ref{sect:over} begins with motivation, focusing on higher types and the idea of non-injective patterns.  We show by example how non-injective patterns could be used in imperative programming including pointer programs.
This idea helps motivate the predicate transformer semantics but is not otherwise developed in this article.
Section~\ref{sect:over} also surveys related work on alternate approaches to programming calculi integrating pointwise with pointfree and functional with imperative styles.

Section~\ref{sect:fun} reviews the standard semantics of simply typed lambda calculus in
a cartesian closed category, in particular $\Mofun$.
Section~\ref{sect:ideal} describes the category of ideal relations,
motivated  by difficulties with semantics in \cite{deMoor:Gib}.  
Section~\ref{sect:rel} gives our relational semantics.
Section~\ref{sect:ilaws} gives a simulation connecting relational and functional semantics, 
and proves the lax beta and eta laws that are our main results for
relational semantics.   
Section~\ref{sect:tran} gives the predicate transformer model and semantics.
Section~\ref{sect:tlaws} proves the main results for transformer semantics.
Section~\ref{sect:disc} assesses the work and discusses open problems.

For Section~\ref{sect:fun} onwards, the reader should be familiar with predicate transformer semantics \cite{Back:book} and with basic category theory including adjunctions and  
cartesian closure \cite{GunterBook}.  
Span constructions and lax adjunctions are only mentioned in passing, and ``laxity'' appears only as an informal term that indicates the weakening of equations to inequations. 

\section{Motivation and background}\label{sect:over}  

\paragraph{Motivation}

One attraction of pointfree style is that it facilitates derivation of programs that are ``polytypic'', i.e., generic in some sense with respect to type constructors~\cite{BackhouseJJM98}.
For example, a polynomial functor on a category of data types may have a fixpoint; its values are trees of some form determined by the particular functor. 
If the element type has a well ordering, one can define the function $repmin$ that sends tree $t$ to the tree $t'$ of the same shape but where each leaf of $t'$ is the minimum of the leaves of $t$.  
De Moor gives a pointfree derivation of $repmin$, at this level of generality, 
using type constructions and equational laws that can be interpreted in functions or in relations~\cite{repmin}.

Relations can model demonic nondeterminacy \cite{deMoor:Gib} or angelic nondeterminacy (as in automata theory and in logic programming), but not both ---unless states or data values are replaced by richer structures such as predicates.
The present author showed that the algebraic structure needed for the polytypic $repmin$ derivation 
exists in the setting of monotonic predicate transformers~\cite{Naumann98a}. 

Although the $repmin$ problem only involves first order data (trees with primitive, ordered data), the derived solution involves higher order: It traverses the input tree to build a closure that, when applied to a value, builds a tree of the same shape with that value at its leaves.  This brings us to a question about how to embed a programming language in a richer calculus for specification and derivation.  In the language of categories, taking data types as objects and programs as arrows, the question is what objects to use for arrow types. 
For each pair $B,C$ of objects, 
a function space
$B\homm C$ exists as an object in the category $\Rel$ of binary relations, and indeed as an object in the category of monotonic predicate transformers.  But $B\homm C$ is not the ``internal hom'' or exponent in $\Rel$. There should be some account of what it means to reason with exponents in $\Rel$ if the derived program is interpreted as a functional one.  

Pointfree reasoning is not without its shortcomings.  De Moor and Gibbons observe that for many specific programming problems a pointwise formulation is easier to understand.  They extend pointwise functional notation to relations by means of non-injective patterns.  
As a simple example, the following is intended to define a relation that performs an arbitrary rotation of a list: 
\begin{equation}\label{eq:rotate}
rotate(x\caten y) = y \caten x 
\end{equation}
An input list  $w$ relates to all $y\caten x$ such that $w=x\caten y$.
Let us consider in detail how a pattern term gives rise to a relation from a set $in$ 
of inputs to a set $out$ of outputs.  The pattern is 
an expression  with free variables $vars$ that are also free in the
result expression.  The situation looks like this:
\[  
\begin{diagram}[h=1.5em,w=3em]
 &           & vars &           & \\
 & \ldTo^{pattern} &   & \rdTo^{expr} & \\
in &          & \rDotsto  &           & out \\
\end{diagram}
\]
As indicated by the dotted arrow,
the semantics of the pattern term is the composition $pattern^o\co expr$.
Throughout this article, $(\co)$ denotes forwards relational composition.

This way of obtaining relations is connected to one systematic approach to embedding programs in richer calculi.  
A \dt{span} in a category is a pair of arrows with common source, like $pattern$ and $expr$ above.  The category $\Rel$ is equivalent to the category whose arrows are spans over the category $\Fun$ of sets and functions.
We refrain from elaborating on the construction but 
note that there are several variations, one of which is a lax span construction that not only gives $\Rel$ from $\Fun$ but also monotonic predicate transformers from $\Rel$~\cite{MartinSCP,Naumann98}.

Let us connect spans with familiar elements of imperative programming.
Consider the humble assignment $x:=x+1$.  
To specify it, beginners often write $x=x+1$ as postcondition, but only miraculous
or divergent programs establish postcondition $\mathit{false}$.
What is needed is an auxiliary variable (``logical constant'') $u$, used in
the specification 
\[ 
pre:\quad x=u \qquad post:\quad x=u+1 
\]
This is interpreted by $u$ being universally quantified over $pre$ and $post$, 
one instance being the useful one that makes $u$ serve to name the initial value of $x$.
Imperative specification notations often feature notation for the special case where the auxiliary is merely equated with an initial value, but the general form is needed, for example to obtain complete laws for sequential composition of specifications~\cite{Morgan:book}.  The general form 
comprises $pre$ and $post$ that are relations between the program state and auxiliary state.
\begin{equation} \label{eq:span_s}
\begin{diagram}[h=1.5em,w=3em]
 &           & aux &           & \\
 & \ldTo^{pre} &   & \rdTo^{post} & \\
in &          & \lDotsto  &           & \mathit{out} \\
\end{diagram}
\end{equation}
The leftward dotted arrow indicates that imperative programs in refinement calculus can be modeled by predicate transformers, i.e., from postconditions to preconditions.  
Still more, the category of monotonic predicate transformers is equivalent to the lax span category over $\Rel$ \cite{MartinSCP}.

One way to view (\ref{eq:span_s}) is that one agent (the angel) chooses the auxiliary value for which the other agent (the demonic program) chooses an output.  Indeed, non-injective patterns can express nondeterministic choice, e.g., the choice between functions $f$ and $g$ can be written 
as $h$ where $h(\mathit{fst}(x,y)) = f\,x ~\mbox{if}~ y~\mbox{else}~g\,x$.
Here $\mathit{fst}$ is the left projection from $A\times\mathit{bool}$ where $A$ is the input type of $f$ and $g$.
In predicate transformer semantics this choice turns out to be angelic, which is consistent with the standard semantics, using $\exists$, of logical constants~\cite{Morgan:book}.

\paragraph{Prior work on the approach explored here}

Although the lax span perspective provides an elegant connection between $\Fun$, $\Rel$, and predicate transformers, it does not directly account for exponents in these categories.  Because we are interested in refinement ---an order relation on arrows--- the exponents, as internal homs, should reflect the ordering.  This led to the investigation of how, by starting from $\Mofun$ instead of $\Fun$, one obtains a more robust category of ``ideal'' relations and then a category ---called $\Tran$ in the sequel--- of monotonic predicate transformers where predicates themselves are monotonic~\cite{Naumann98}. 
One step towards a programming calculus based on $\Tran$ is to show~\cite{Naumann98b} 
that (lax) polynomial functors have unique fixpoints; 
the lax exponent serves to derive a variation on Lawvere's parameterized recursion theorem. 
Further steps are taken in \cite{Naumann98a} 
which adds the pointfree theory of containers/membership \cite{OegeMembJ} and uses that, together with the recursion theorem, to recreate a derivation of de Moor's \cite{repmin} pointfree $repmin$ solution at the level of $\Tran$.
What is still not achieved after all these years is to derive a solution that uses shared pointer structure:
updating all leaves of the tree to point to a single cell that is updated with the minimum of the tree, once that has been determined.

\paragraph{Towards refinement calculus for pointer programs}

Extant refinement calculi for imperative programming do not address pointer programs except by explicit encodings of the heap using arrays.
For \emph{post hoc} program verification there has been a great deal of progress using separation logic.\footnote{A recent reference that emphasizes separation algebra is the book by Appel et al \cite{AppelCertComp}.}
The assertion language features the separating conjunction, $*$, and assertions are often in the form of a top level separation $P*Q*\ldots *R$ that expresses some way of partitioning heap cells into disjoint sets that satisfy the predicates $P,Q,\ldots$.  Informally, the sets are referred to as ``footprints'' of predicates.  Some static analysis tools 
use separated conjunctions as ``symbolic heaps'' in symbolic execution.
Underlying the semantics of the assertion language is a notion of separation algebra, whose expressions denote (partial) heaps and whose operations manipulate heaps.
However, for reasoning about programs, separation logic distinguishes programs from specifications rather than embedding one in the other as in refinement calculus.  And reasoning must be done in terms of assertions; put differently, in terms of expressions only of type `proposition'.   

One of the challenges in reasoning about pointer programs 
is to describe structural invariants such as reachability, uniqueness of references, separation, and confinement.
Separation facilitates reasoning about independence of writes, i.e., frame conditions.
This is embodied in the frame rule of separation logic:
 \begin{equation}\label{eq:frame}
\mbox{from} \quad \{P\}\, cmd\, \{Q\} \quad \mbox{infer} \quad \{P*R\}\,cmd\,\{Q*R\} \; .
 \end{equation}
This is sound because the footprint of precondition $P$ serves as frame condition for the command:
the antecedent implies that $cmd$ does not write outside that part of the heap.
The precondition $P*R$ says that $R$ is true of a disjoint set of heap cells, so it cannot be falsified by writes to the cells that support $P$.

By contrast with separation, confinement facilitates reasoning about independence of reads.
One application is in reasoning about simulations between data representations: 
Confinement helps ensure that the behavior of a client of an abstract data type is independent from the internal representation because the client has no pointer into the representation.  
Wang, Barbosa, and Oliveira \cite{WangBO08} extend separation logic with connectives that express confinement.
For example, $P \neg\!\triangleright Q$ says that not only are the footprints of $P$ and $Q$ disjoint but no cell in the footprint of $P$ has a pointer to a cell in the footprint of $Q$.
This is an elegant advance on works which use ad hoc means to express confinement (e.g., \cite{BanerjeeNaumann02c,banerjeeN13}). 
Dang and M\"{o}ller~\cite{DangMoeller15} take further steps in this direction.
 
Another challenge in reasoning about pointer programs is that programming languages are impoverished in their means to express heap operations ---it amounts to little more than load and store instructions.  
Separation logic was a huge advance in reasoning about such programs, which involves assertions about many intermediate states in which interesting invariants are temporarily broken.    
Informally, what works effectively is diagrams.  

Consider, for example, 
an iterative algorithm that reverses, \emph{in situ}, an acyclic doubly linked list.
The algorithm uses pointer variable $p$ pointing to the reversed segment and $n$ that 
points to the first node of the segment that remains to be reversed.  
The loop body makes three heap updates and moves $p$ and $n$ forward,
transforming the left-hand situation into the right-hand one.  
\begin{center}
\includegraphics[width=4.5in]{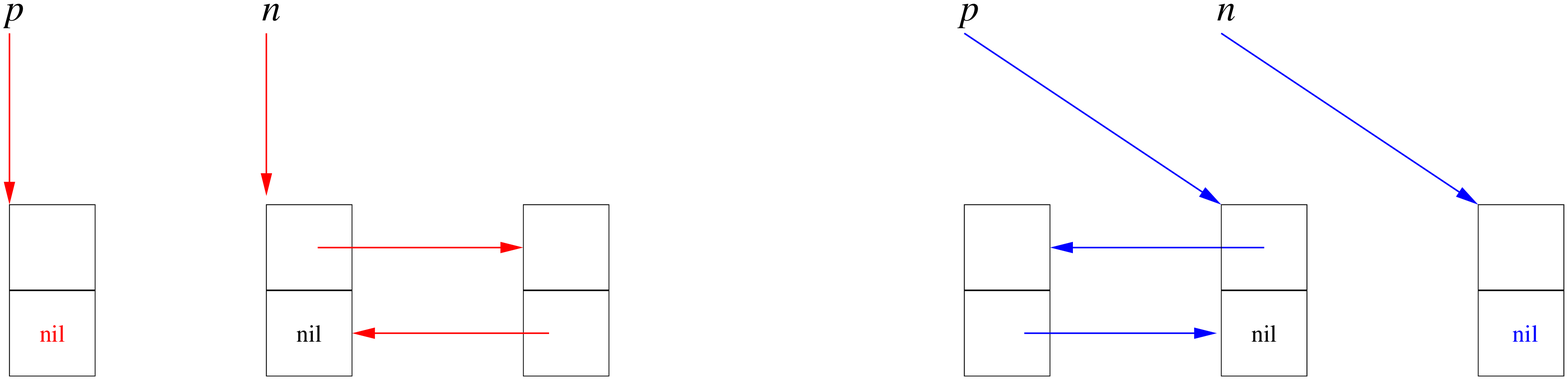}   
\end{center} 
Each rectangle represents a cell with two pointers, which in a high level language could be fields
named $next$ (upper box) and $prev$ (lower) for the forward and backward pointers. 
Here is another diagram that depicts the transformation.\footnote{Diagrams can mislead, of course.  In the general case, the rightmost and leftmost pointers, in and out of the clouds that indicates the rest of the heap, may not exist.}
\begin{center}
\includegraphics[width=4.4in]{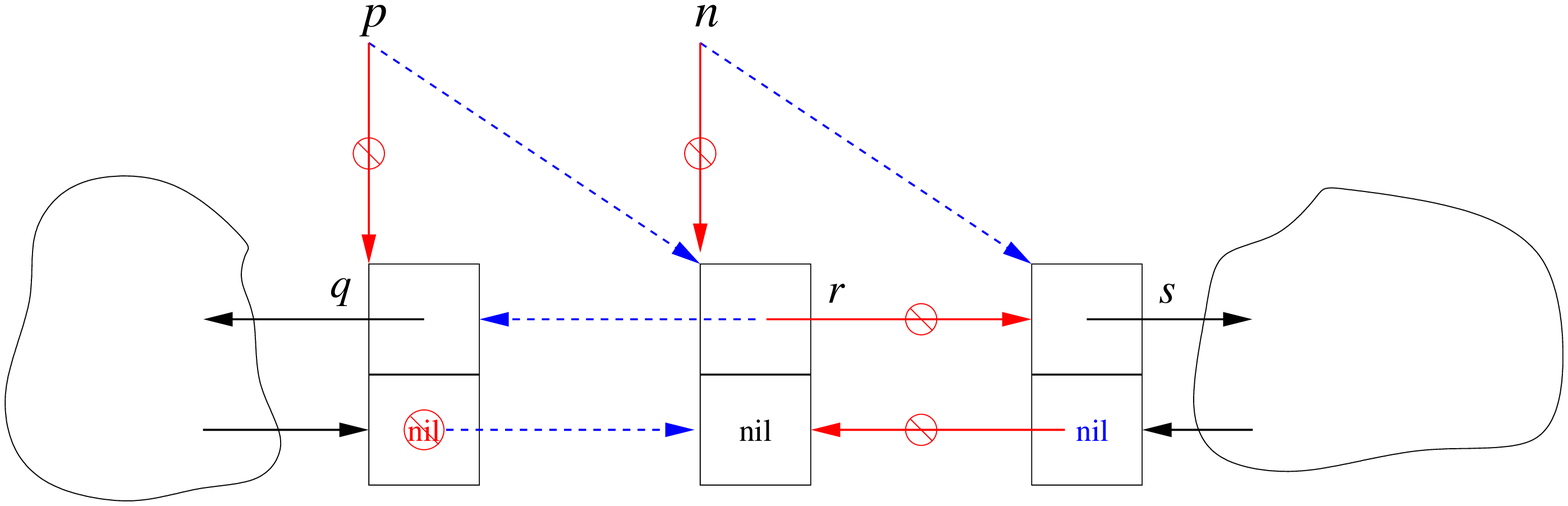}   
\end{center} 
Dashed arrows indicate the {\color{blue}final state}.
Slashed circles indicate values in the {\color{red}initial state} that get updated.
For example, the left cell's $prev$ is initially nil and the right cell's $prev$ gets set to nil.

The following notional code is a tail-recursive procedure $rev1$ that uses pattern matching to perform the transformation in a single step.  
\[\begin{array}{lcl@{\hspace*{.7em}}l@{}l@{}l@{}l@{}l}
rev1\ p\ n\ heap &\mathord{=}& 
\mathbf{match}\ heap & \!\!\mathbf{with} \\
&& \mid & \cell{p}{q}{\nil} &\Sep& \multicolumn{3}{@{}l}{\cell{n}{\nil}{\nil} \quad \BigPatarrow} \\
&&      & \cell{p}{q}{n}    &\Sep& \cell{n}{p}{\nil} \\
&& \mid & \cell{p}{q}{{\color{red}\nil}} 
                            &\SEP& \cell{n}{{\color{red}r}}{\nil} 
                            &\SEP& \cell{r}{s}{{\color{red}n}} \quad\; \BigPatarrow \\
&&\quad\:rev1\ n\ r\ ( \!\!\!\! & \cell{p}{q}{{\color{blue}n}} 
                            &\SEP& \cell{n}{{\color{blue}p}}{\nil} 
                            &\SEP& \cell{r}{s}{{\color{blue}\nil}} \; ) 
\end{array}
\]
The second pattern clause matches the diagram.
(The first clause handles the situation where $n$ points to the last node.)
As usual in pattern matching,
free variables on the left of the pattern arrow $\BigPatarrow$ are bound by the match.
That is indicated in the diagram by labels $q,r,s$.

The notation is adapted from separation logic and from OCaml.
The difference from OCaml's match construct is the big arrow $\BigPatarrow$ chosen to fit with other notations in this article.
The expression $\cell{x}{y}{z}$ is inspired by the points-to predicate in separation logic. 
It denotes a heap with a single cell, referenced by $x$,
with $y,z$ as the values of the $next$ and $prev$ fields.  
Think of the heap as a mapping from references to pairs, so 
$\cell{x}{y}{z}$ simply denotes a singleton mapping. 
Infix $*$ denotes the partial function that forms the union of two heaps, provided their domains are disjoint, and is undefined otherwise.
Comma, for pairing, binds tighter than $\cell{}{}{}$ which in turn binds tighter than $*$.

The initial call to $rev1$ should be $rev1\ root\ n\ heap$ where $root$ points to the first node
and $n$ is that node's $next$ (its $prev$ being $\nil$).  

One wishful feature of the code for $rev1$ is that the patterns are local, in the sense that they describe only the relevant cells.
Given that the heap is an explicit argument and return value, 
perhaps the pattern should match the entire heap. 
To see how that might look, consider 
the following putative definition of $reverse$, which makes the top level call to $rev1$ for lists of plural length.
\[\begin{array}{lcl}
reverse\ \nil\ hp &=& hp \\
reverse\ p\ (\cell{p}{\nil}{\nil} \Sep hp) &=& (\cell{p}{\nil}{\nil} \Sep hp) \\
reverse\ p\ (\cell{p}{n}{\nil} \Sep hp) &=& rev1\ p\ n\ (\cell{p}{n}{\nil} \Sep hp) 
\end{array}
\]
It would be nicer for $hp$ to be implicit, by analogy with the frame rule (\ref{eq:frame}).
An obvious approach would be to use a state monad.
Another possibility is to find algebraic notation and laws whereby a whole-heap operation can be derived from one written using a local-heap pattern term.  
Laws are also needed to transform heap patterns into some restricted form that can be 
automatically compiled to code.

In separation algebra the heap union operator, $*$, is partial.
For example, $\cell{x}{y}{z} * \cell{x}{u}{w}$ is not defined.
Our wishful pattern notation needs to be interpreted in a setting that admits patterns that are non-total as well as non-injective.
The calculus of de Moor and Gibbons allows general relations in patterns ---but there are unsettled issues concerning refinement monotonicity at higher types, as discussed in Section~\ref{sect:ideal}.  
In our relational model, the issues are addressed by restricting pattern terms to total functions.  
The predicate transformer model supports relational pattern terms.

\paragraph{Related work on other approaches}

It does not appear that the specific calculus of de Moor and Gibbons has been developed beyond the original paper \cite{deMoor:Gib}.

Hinze makes nice use of explicit powersets as a way to do pointwise relational programming~\cite{Hinze02}.


Bunkenburg~\cite{BunkenburgDiss} develops a calculus of  
``expression refinement'', for derivation of lazy functional programs including those that use the state monad.
Ordering is taken into account by treating data types as
ordered sets.  Nondeterministic expressions are interpreted as monotonic functions 
$\mor{f}{A}{\U B}$ where $\U B$ is the set of \dt{updeals} (upward closed subsets) on $B$.
The set $\U B$ is ordered by $\sps$, which coincides with the Smyth order
(total correctness) restricted to upward closed sets.  Upward closure is natural for
outcomes of expressions.
Monotonic functions $\mor{f}{A}{\U B}$ are naturally isomorphic to ideals, as is
discussed in Section~\ref{sect:ideal}.
So there is some overlap with the present work, 
although our language (and~\cite{deMoor:Gib}) is by-value. 
Bunkenburg develops a wide spectrum language and a logic of refinement, but the language does not include pattern matching.

Uses of angelic and demonic nondeterminacy are explored by Back and von Wright~\cite{Back:book} and by Morris and Tyrrell~\cite{MorrisT08toplas}.  
Martin, Curtis, and Rewitzky~\cite{MartinCR07} use multi-valued relations as an alternative to predicate transformers for modelling the combination of angelic and demonic nondeterminacy,
and adapt map/fold algebra to that model \cite{MartinC08}.

Morris, Bunkenberg, and Tyrrell~\cite{MorrisBT09}
introduce a novel ``phrase'' construct, and ``term transformer'' semantics, as a technique for specifying and reasoning about imperative programs; it generalizes predicate transformers to terms of any type.
Their approach encompasses unbounded angelic and demonic nondeterminacy
together with higher order functions, for which Morris and Tyrrell present numerous 
laws~\cite{MorrisT08toplas}.   
They sketch how non-injective patterns may be reduced to nondeterminacy.  
They have beta-reduction as an equality, unlike our results in the sequel, 
owing to partitioning terms as proper or not, and use of a semantic monotonicity restriction.
Morris and Tyrrell  have also shown equivalence between monotonic predicate transformers, multirelations, and free completely distributive lattices~\cite{MorrisT08acta}.
The phrase construct is very appealing but quite different from the functional notations explored in this article,
and from conventional notations of refinement calculi.


For first order concurrent programs, concurrent Kleene algebra provides a pointfree algebraic setting for reasoning about separation, mutable state and concurrency (e.g.,~\cite{HoareSMSVZO14}).  
Modal Kleene algebra is the basis for the extended separation logic of Dang and M\"{o}ller~\cite{DangMoeller15},
which they use to verify algorithms including list reversal and tree rotation, expressed in terms of  load and store commands.
Banerjee et al~\cite{RegLogJrnI} use explicit expressions to designate footprints of pointer structures, but only for \emph{post hoc} verification.

\section{Functional semantics}\label{sect:fun}

This section reviews the categorical semantics of simply-typed lambda calculus
in the order-enriched category $\Mofun$ of monotonic functions on (small) posets.

An \dt{order-enriched category} is one with homsets partially ordered and composition
monotonic.  
Each  homset $\Mofun(B,C)$ carries
the pointwise ordering  $f\pol g \equ (\all b \bdot fb\pol_C gb)$.  
Composition of $f\in\Mofun(B,C)$ with $h\in\Mofun(C,D)$, written $(f\co h)$ 
as with relations in general, is monotonic in both $f$ and $h$.  
We write $\times$ for binary product and 
$\mor{\LL f,g\RR }{B}{C\times D}$ for pairing of $\mor{f}{B}{C}$ with $\mor{g}{B}{D}$.
For projection we write
$\pi$ with subscripts or some other indication of which projection is meant.
We write $B\homm C$ for function space as an exponent object of $\Mofun$, ordered
pointwise.  Let $\homm$ bind less tightly than $\times$. The application function is
denoted by $\mor{\ap}{(B\homm C)\times B}{C}$ and currying sends $\mor{f}{B\times C}{D}$ to 
$\mor{\cur f}{B}{C\homm D}$.  The decorative subscript ``\textsf{f}''
distinguishes the functional constructs from those for ideals (\textsf{i}) and 
predicate transformers (\textsf{t}) to come.

\begin{proposition}
\label{ax:mo}
\upshape
$\Mofun$ is cartesian closed and the adjunctions $(B\times)\dashv(\homm B)$ are
order-enriched.  
\end{proposition}
Order enrichment, in the context of Proposition~\ref{ax:mo}, simply means that pairing, currying, and the exponent and product
functors are monotonic.

The \dt{functional terms} $M$ are those of simply-typed lambda calculus.
Types are given by 
\begin{equation}\label{eq:types}
\typ::= b \mid \one \mid  \typ \times \typ  \mid  \typ \typarrow \typ 
\end{equation}
where $b$ ranges over some given set of base types. Terms are generated from given constants $c$ and variables $x$:
\begin{equation}\label{eq:funcGram}
 M ::= x \mid  c \mid \mathit{fst} \mid  snd  \mid \LL M,M\RR \mid  M M 
   \mid \lambda x.M   
\end{equation}
Typing judgements take the form $\Gamma\tg M:\typ $ where $\Gamma$ is a list 
$x_0:\typ _0,\ldots,x_n:\typ _n$ of variable typings. The  typing rules are 
standard and omitted \cite{GunterBook}.  We allow constants at all types. The type of a constant is assumed given.

We write $\F\se\slot\es$ for the functional semantics, but omit $\F$ in this section. 
Functional semantics is based on a given poset $\Fi\se b\es$ for each base type $b$.
Given a singleton set $\se\one\es$ as well, the semantics of types is given inductively
by $\se \typ \times \typ '\es = \se \typ \es \times \se \typ '\es$ and 
$\se \typ \typarrow \typ '\es = \se \typ \es \homm \se \typ '\es$.
For  $\G=x_0:\typ _0,\ldots,x_n:\typ _n$, we 
 write $\se\G\es$ for the product 
$(\ldots((\se\one\es\times\Fi\se \typ _0\es)\times \Fi\se \typ _{1}\es)\times\ldots \Fi\se \typ _n\es)$.  
For terms, we assume that for each constant $c:\typ $ a function 
$\mor{\Fi\se c\es}{\se\one\es}{\Fi\se \typ \es}$ is given, and let 
$\mor{\Fi\se\mathit{fst}\es}{\se\one\es}{\Fi\se \typ \times \typ '\typarrow \typ  \es}$  
pick out the left projection function (as an element of the set 
$\Fi\se \typ \times \typ '\typarrow \typ  \es$).
The semantics of $\G\tg M:\typ $ is a 
morphism $ \mor{\Fi\se \G\tg M:\typ \es }{\Fi\se\G\es}{ \Fi\se \typ \es} $,
defined in a standard way in Table~\ref{tab:fsem}.\footnote{See, for example, \cite{GunterBook}. Our formulations use only binary products. The ``appropriate projection'' for $x_k$ is the evident morphism
$\se\G\es \arrow \se \typ_k\es$ defined using binary projections and pairing.  
}

\begin{table}[t]
    \caption{The functional semantics $\F\se\slot\es$,
given functions $\se c:\typ \es: \one \arrow \F\se\typ\es$.
}
    \label{tab:fsem}
\[
\begin{array}{lcl}
\Fi\se \G\tg x:\typ\es & = & \pi \quad\mbox{where $\pi$ is the appropriate projection} \\
\Fi\se \G\tg c:\typ\es & = & \pi\co\Fi\se c:\typ\es \quad\mbox{where $\pi$ is the projection
 $\se\G\es\arrow\se\one\es$} \\ 
\Fi\se \G\tg \LL M,M'\RR:\typ \times \typ ' \es & = & \LL \Fi\se \G\tg M:\typ \es, \Fi\se \G\tg
            M':\typ '\es\RR \\
\Fi\se \G\tg M M':\typ \es & = & \LL \Fi\se \G\tg M:\typ '\typarrow \typ \es, \Fi\se\G\tg
M':\typ '\es\RR \co \ap \\ 
\Fi\se \G\tg \lambda x. M:\typ \typarrow \typ '\es & = &  \cur\Fi\se \G,x:\typ \tg M:\typ '\es 
\end{array} 
\]
\end{table}

\section{Ideals}\label{sect:ideal}

This section motivates and defines the category $\Idl$ of ideal relations, 
starting from a sketch of difficulties that arise in the
work of De Moor and Gibbons \cite{deMoor:Gib} using the category $\Rel$ of relations on sets.
We describe the embedding of $\Mofun$ in $\Idl$, the power adjunction which
gives an embedding the other way around, and the lax cartesian closed structure used in
Section~\ref{sect:rel} for semantics.
Results that are not proved here can be found in \cite{Naumann98} or \cite{Naumann01b}. 

The relational semantics of \cite{deMoor:Gib} interprets arrow types using the set 
$A\homi B$ of relations from $A$ to $B$.  
For relation $\mor{R}{B\times C}{D}$ the pre-curried relation
$\mor{cur~R}{B}{C\homi D}$ is defined by 
\begin{equation}\label{eq:cdef}
 b(cur~ R)S \;\equ\; (\all c,d \bdot cSd \equiv (b,c)Rd) 
\end{equation} 
But $cur$ is not monotonic with respect to $\sps$, and monotonicity with respect to
refinement is crucial for a useful calculus and for recursion.
De Moor and Gibbons address monotonicity by replacing $\sps$ with a refinement order defined by
$ R\pol R' \; \equ \; R\co\rsps \ssps R' $
for $R,R'$ of type $B \arrow C\homi D$.  
(Small font is used when order relations are composed.)
In terms of points: $ R\pol R'$ iff for all $b,S$, $bR'S$ implies there is $S'$ with $bRS'$ and $S'\sps S$.
At higher types, refinement is defined inductively:
\begin{equation}\label{eq:dMo}
R\pol R' \; \equ\;  R\co\rpol \ssps R'
\end{equation}
where the small $\rpol$ is the refinement order on the target type.  
Observe that $\pol$ reduces to $\sps$ 
for relations such that $R\co\rpol = R$. 
The ordering has a problem:  $(Q\co R)$ fails to be monotonic in $Q$, e.g., if
$\mor{R}{C\homi   D}{C\homi D}$ sends a relation to its complement.  
Our solution is to rule out complementation on grounds of
non-monotonicity.
A function $\mor{f}{B}{C}$ is monotonic iff 
$f\co\rpol_C \sps \rpol_B \co f$.
Let us say a relation is \dt{monotonic} iff 
$ R\co\rpol \ssps \rpol\co R$.  
It is straightforward to show that $(\co)$ is monotonic,
w.r.t.\ $\pol$ of (\ref{eq:dMo}), when restricted to monotonic relations.\footnote{The symbols $\ss$ and $\sps$ always have their usual set-theoretic  meaning, whereas $\pol$ is used
generically for orderings.  We write $id$ only for identity functions,  writing either
$\Id$ \emph{sans serif} or $\pol$ for the identity in $\Idl$, depending on which seems
most perspicuous. }

If  $R$ is monotonic then so is $R\co\rpol$.
A relation is monotonic  and satisfies $R\co\rpol = R$ iff it is an ideal.
An \dt{ideal} is a relation $\mor{R}{A}{B}$ 
such that 
\[\rpol_A \co R\co \rpol_B \quad\mathrel{\ss}\quad R 
\]
In terms of points: $a'\pol a$, $aRb$, and $b\pol b'$ imply $a'Rb'$.  

Ideals are the morphisms of the order-enriched category $\Idl$ whose objects are all
(small) 
posets. Composition $(\co)$ in $\Idl$  is the same as in $\Rel$, but the identity on
$A$, written $\Id$, is $\pol_A$. Homsets are ordered by $\sps$, not $\ss$,
for reasons mentioned later.


\paragraph{Ideals, power adjunction, and products}

A \dt{comap} in an order-enriched category is an arrow $g$ with a corresponding \dt{map}
or \dt{left adjoint} $g^*$. That is, $g$ and $g^*$ satisfy 
\begin{equation}\label{eq:adjoints}
\Id\pol g^*\co g \quad \mbox{and} \quad g\co g^*\pol \Id 
\end{equation}
It is common to order $\Rel$ by $\ss$ and describe  $\Fun$ as the subcategory of  maps of
$\Rel$; then the comap of a function $f$ is its converse $f^o$.
Maps in order-enriched category $\cat{C}$ are the comaps of the arrow-dual $\cat{C}^{op}$ and
they are also comaps in the order-dual $\cat{C}^{co}$ obtained by reversing
the order on homsets. 
As $\Rel$ is isomorphic to both $\Rel^{op}$ and $\Rel^{co}$,
one has several opportunities for making infelicitous choices of nomenclature.
Our choices smooth some parts of the exposition, but have the unfortunate consequence 
that functions embed as comaps in $\Idl$.
The reader should keep in mind that, in this article, the operation $^*$ gives the map for a comap, rather than the reverse.
For example, in $\Idl$ we have the shunting property
\begin{equation}\label{eq:shunt}
c\co R \sps S \quad\mbox{iff}\quad R \sps c^* \co S \quad \mbox{ for comap } c
\end{equation}
which is a standard consequence of (\ref{eq:adjoints}), 
instantiating $\pol$ with the order $\sps$ on homsets of $\Idl$.

Because functions are a special case of relations, 
$\Fun$ is included in $\Rel$.
But monotonic functions are not a special case of ideal relations.  
If $f$ is in $\Mofun(A,B)$ and $R$ is in $\Idl(B,C)$ then $(f\co R)$ is in
$\Idl(A,C)$.  As a consequence, there is a 
\dt{graph functor} $\mor{\Gr}{\Mofun}{\Idl}$ defined by $\Gr f = (f\co\rpol)$ and
$\Gr A = A$. We can sometimes elide  $\Gr$ because 
\begin{equation}\label{eq:gr} \Gr f \co R = f \co R \quad \mbox{for ideal $R$}. 
\end{equation}
The graph functor is an order injection on homsets, because
the pointwise ordering of functions $f\pol g$ is equivalent to the inclusion
$(f\co\rpol)\sps (g\co\rpol)$.

An unfortunate feature of $\Idl$ as compared with $\Rel$ is that the converse 
of an ideal $\mor{R}{B}{C}$ need not be an ideal of type $C\arrow B$ unless
$B$ and $C$ are discretely ordered.  
Note that $\mor{R^o}{\op{C}}{\op{B}}$ is an ideal, where 
$\op{B}$ is the order dual of $B$, i.e., $\pol_{\op{B}}$ is $(\pol_{B})^o$.

More importantly, for $\mor{f}{B}{C}$ in $\Mofun$, 
$\mor{(\rpol\co f^o)}{C}{B}$ is an ideal; in  fact it is the left adjoint of $\Gr f$.
The \dt{opgraph functor} $\Rg$ is defined as 
\[ \mor{\mathsf{Rg}}{\Mofun^{co\,op}}{\Idl}
\qquad \mathsf{Rg}\,f = (\Gr f)^* \] 
So we have $\Id\sps \Rg\,f\co\Gr\, f$ and 
$\Gr\,f \co \Rg\,f\sps \Id$.

A functor $G$ on order-enriched categories is an \dt{embedding} if it is an
order injection on homsets, i.e., $f\pol h\equ G f\pol G h$ for all
$f,h$, and  it is bijective on objects.  

\begin{proposition}
\label{ax:emb}
\upshape 
$\Gr$ embeds $\Mofun$ onto the comaps of $\Idl$.
\end{proposition}

Recall from Section~\ref{sect:over} that $\U B$ is the set of updeals, i.e., upward closed subsets, of $B$, ordered by $\sps$.  
We use the power adjunction \cite{Freyd:Scedrov} to characterize 
updeal lattices $\U B$ in $\Idl$.  
The converse $\Ni$ of the membership relation is an ideal $\U B\arrow B$;  
note that $(\Ni\co\rpol_B) \ss \mathop{\Ni}$ says that $\U B$ contains only updeals.
For $R$ in $\Idl(B,C)$, the function $\mor{\Lambda R}{B}{\U C}$
sends $b$ to its direct image  through $R$ (i.e., the set of $c$ such that $bRc$).  
The monotonic functor
$\mor{\U}{\Idl}{\Mofun}$ is defined on morphisms by $\U R = \Lambda (\Ni\co R)$.
\begin{proposition} 
\upshape 
$f = \Lambda R  \equ  f\co\Ni = R$, for all $R,S$ in $\Idl$ and  $f$ in $\Mofun$.
This is order-enriched:
$ R\sps S\equ\Lambda R \pol \Lambda S $,  where
$\pol$ is the pointwise order on functions, lifted from the order $\sps$ on
$\U C$.  
Furthermore, $\U$ is an  embedding of $\Idl$ in $\Mofun$,
and $\Gr$ is left adjoint to $\U$ with counit $\Ni$ and unit 
the upward closure. 
This is order enriched: $f\pol g \equ \Gr f \sps \Gr g$. 
\end{proposition}
Owing to our choice of  ordering on $\Idl$, the power
adjunction is order-enriched (by contrast with \cite{BirdDeM}),   
and covariantly so (by contrast with \cite{Naumann98}).

For the product $A\times B$ of posets, the projection function
$\pi$ in $ \Mofun(A\times B,A)$ gives a comap $\Gr \pi$
in $ \Idl(A\times B,A)$.
We overload notation and write  
$\mor{\LL R, S\RR}{D}{A\times B}$
for the pairing of $\mor{R}{D}{A}$ with $\mor{S}{D}{B}$, which 
is defined by 
$ d\LL R, S\RR(a,b) \equ d R a \land d S b $ just as in $\Rel$.  
Clearly $\LL\Gr f,\Gr g\RR = \Gr \LL f, g\RR$.
Defining $Q\times R$ as usual makes $\times$ an extension of the product
functor on $\Mofun$ in the sense that 
$ \Gr f\times \Gr g = \Gr(f\times g) $.
\begin{proposition}
\upshape 
$\times$ is a monotonic bifunctor on $\Idl$ and 
\begin{eqnarray}
\label{eq:pra}
R &\sps & \LL R, S\RR \co \Gr \pi  
\quad(=, \mbox{ if $S$ is a comap})  \\
\label{eq:praa}
S &\sps & \LL R, S\RR \co \Gr \pi'  
\quad(=, \mbox{ if $R$ is a comap})  \\
\LL R\co\Gr \pi \,,\, R\co\Gr \pi'\RR &\sps& R \quad(=, \mbox{ if $R$  comap}) 
\end{eqnarray}
\end{proposition}
Here $\pi$ and $\pi'$ are the left and right projections.
We let composition $(\co)$ bind more tightly than the pairing comma, as in these
laws used later:\footnote{In the categorical lingo, $\times$  is  locally
  right adjoint to the diagonal $\mor{\Delta}{\Idl}{\Idl^2}$.
  It is an adjunction on comaps and (by general abstract nonsense \cite{pre:adj}) 
  it is the unique (up to isomorphism) local right adjoint to the diagonal that is an
  adjunction on comaps.  
  In this article  we consider lax exponents of relations and lax products/exponents for
   predicate  transformers.  Each can be
  characterized up to isomorphism by inequations  (conditional in some cases) together with 
  equations conditioned on certain morphisms being maps, comaps, or both (see \cite{Naumann98}).  
  In this article we need properties of the structures but not their uniqueness.
} 
\begin{eqnarray}
\label{eq:prb}  
 \LL R\co S \,,\, R\co U \RR & \sps & R\co\LL S,U\RR \quad(=,\mbox{ if $R$ comap}) \\
\label{eq:prc}  \LL R,S\RR\co (U\times V) & = & \LL R\co U \,,\, S\co V \RR
\end{eqnarray}

\paragraph{Relation space}

In the same way that $\times$ extends to a functor on $\Idl$, the sum and
function-space constructs can be extended.  But the function space is not
an internal hom of $\Idl$, which is what we need to interpret a lambda calculus.
We now construct currying and application for the exponent defined by $B\homi C = \Idl(B,C)$.
The impatient reader may care to skip to Prop.~\ref{ax:ex} which summarizes what is needed in
the  sequel.

In $\Rel$, an exponent is given by the adjunctions $(\times B)\dashv(B\times)$.
Although $\times$ does not have this property in $\Idl$,  there is a related
functor denoted by $\ltimes$ that gives $(\times B)\dashv(B\ltimes)$ in $\Idl$.
Formally, this gives an interpretation of lambda terms, but a strange one
because it does not extend $\homm$.
We use $\ltimes$ only as a stepping stone.  The currying operation $\curs$
associated with $(\times B)\dashv(B\times)$ has the effect 
of shifting one coordinate.  It is defined for
$R\in\Rel(A\times B,C)$ just like (\ref{eq:cdef}), that is:
\begin{equation}\label{scdef} a (\curs  R)(b,c) \equ (a,b) R c \end{equation}
This defines $\curs R$ as a relation in $\Rel(A,B\times C)$ but it need not be
an ideal even if $R$ is, because the pointwise order on
$B\times C$ is incompatible with $R$ being downward closed in $B$.
So we define $B\ltimes C$ to be the poset $\op{B}\times C$, i.e.,
the set $B\times C$ but with the 
left-contravariant order ${\pol_{B\ltimes C}} = {(\pog_B\times\pol_C)}$.  Now $\curs R$
defined by (\ref{scdef}) is an ideal if $R$ is. 
The application ideal $\mor{\aps}{(B\homs C)\times B}{C}$ has 
$ ((b,c),b') \, \aps\, c' \equ b\pog_B b' \land c\pol_C c' $.
It is straightforward to prove that $\curs$ is monotonic, and that it has an inverse.
For each $B$, the functor $(B\ltimes)$ is right adjoint to $(\times B)$.

By analogy with relations as subsets of a cartesian product, we define $\homi$ by
\begin{equation}\label{eq:hd} B\homi C = \U(B\homs C) \end{equation}
An upward closed subset of $B\homs C$ is, 
by definition of the order, an ideal,  
so $B\homi C$ is another name for the poset $\Idl(B,C)$. 
For ideals $\mor{R}{B\times C}{D}$ and  $\mor{S}{B}{C\homi D}$ we define
$\mor{\curi R}{B}{C\homi D}$ and $\mor{\uncuri S}{B\times C}{D}$ by
\[ \curi R = \Gr(\Lam (\curs R))  \qquad\quad
\uncuri S = \uncurs(S\co\Ni) 
\]   
Observe that $\Lam (\curs R)$ is the monotonic function sending $b\in B$ to
the updeal $\{ (c,d) \mid  (b,c)R d\}$,
which is essentially ``$R$ curried on $b$'' (cf.\ (\ref{eq:cdef})).
Monotonicity here means that $b\pol b' $ implies $\Lam(\curs R) b \ssps \Lam(\curs R) b'$.

By definition of $\Gr$ and the ordering $\sps$ on $C\homi D$, an element $b$ is related by
$\curi R$ to all ideals contained in $\{ (c,d) \mid  (b,c)R d\}$.
Clearly $\curi R$ is monotonic in $R$, as $\Lam$ and $\curs$ are monotonic.
A crucial fact about $\curi R$ is that it is a comap, for any $R$.  This follows by
definition of $\curi$ and Prop.~\ref{ax:emb}.  
We also have 
\( 
\uncuri (\curi R) = R 
\)
by a straightforward calculation using power and $\ltimes$ adjunctions.
For $\curi(\uncuri S)$, however, unfolding the definitions and using $\ltimes$
adjunction yields $\curi(\uncuri S) = \Gr(\Lam(S\co \Ni))$
and then power adjunction gives 
\begin{equation} \curi(\uncuri S) = S \quad\mbox{if $S$ is a comap}\end{equation}
The condition is necessary; 
the range of $\curi$ is comaps so it is not surjective.
Unfolding the definitions at the level of points, 
we have $(b,c)(\uncuri S)d $ iff $(\some R \bdot bSR \land cRd) $, 
and thus
$ b(\curi(\uncuri S))U $ iff
$ \{ (c,d) \mid (\some R \bdot bSR \land cRd) \} \sps U $.
The largest such $U$ is thus a ``convexification'' of $S$: For given $b$, one can say 
that $S$ nondeterministically chooses a $V$ which for given $c$ nondeterministically
chooses result $d$; whereas the largest $U$ above combines all the nondeterministic
outcomes of all $V$.  
So $\curi(\uncuri S)$ is $\Gr f$ for the function $f$ such that 
$f~b$ is the convex closure of $S$ for $b$.  
From this it is clear that 
\begin{equation} \label{eq:curi-uncuri}
\curi(\uncuri S) \sps S \quad\mbox{for all } S\end{equation}
Defining $\api = \uncuri\,\Id$ makes $\api$  an ideal with
$ (R,b)\api c \equ bRc $.  
(It is not a comap.)
Note that $\uncuri\,\Id = \uncurs \Ni$ with  $\Ni$  taken to be an ideal 
$(B\homi C)\arrow B\ltimes C$ (recall (\ref{eq:hd})).
Application extends that for functions, in the sense that 
\[ (\Gr f,a)\api\, b \quad\mbox{iff}\quad (f,a)(\Gr\, \ap)b\quad\mbox{iff}\quad
fa\pol b  \]
We have $\uncuri S = (S\times \Id)\co\api$ (recall that $\Id$ is $\pol$).  
For $\mor{R}{A}{B}$ and $\mor{S}{C}{D}$, we define
$\mor{R\homi S}{B\homi C}{A\homi D}$ by
\[ R\homi S = \curi((\Id_{B\homi C}\times R)\co\api\co S) \]
as usual.
This extends $\homm$ because $U(R\homi S)V \equ R\co U\co S \sps V$.  
Note that $R\homi S$ is a comap for all $R,S$, because the range of $\curi$ is comaps. 

\begin{proposition}
\label{ax:ex} 
\upshape 
$\mor{\homi }{\Idl^{op}\times\Idl}{\Idl}$ is a monotonic functor, and 
for all $R,S,U$ 
\begin{eqnarray}
\curi R \mbox{ is a comap} & &\\
\label{eq:ap}
(\curi R \times \Id )\co \api &=& R\\  
\label{eq:apx}
\curi((S\times \Id)\co \api) &\sps&  S \quad\mbox{($=$, if $S$  comap)}\\
\label{eq:curnat}
\curi((R\times \Id )\co S) & \sps & R\co \curi S \;\mbox{($=$, if $R$ comap)} \\
\curi((\Id\times S)\co R\co U)&=&\curi R\co (S\homi U)
\end{eqnarray}
\end{proposition}

\paragraph{Internalizations}

To connect $B\homm C$ with $B\homi C$, we define the ideal $\gr_{B,C} $ by 
\[ \mor{\gr_{B,C}  = \curi(\Gr\,\ap)}{(B\homm C)}{(B\homi C)} \]
This makes $\gr_{B,C}$  the internalization of the action of $\Gr$ on a
homset: it relates each $f$ in $B\homm C$ to the set of ideals contained in $\Gr f$. 
The following expresses how $\curi$ and $\api$ extend their functional
counterparts when applied to comaps.
\begin{equation}
\label{eq:emac}
\curi (\Gr f) = \Gr(\cur f) \co \gr  \qquad\quad
(\gr \times\Id)\co \api = \Gr\,\ap    
\end{equation}
For any $B,C,D$, there is an ideal
$\mor{\comp}{(B\homi C)\times(C\homi D)}{B\homi D}$ which internalizes composition.  
It is defined in a standard way using $\curi$, $\api$, and structural isomorphisms like 
associativity for products (Prop.~\ref{ax:mo} and Prop.~\ref{ax:emb}).

\section{Relational semantics}\label{sect:rel}

The language of \dt{relational terms} $N$ has the same types as before, 
see (\ref{eq:types}).
But in addition to functional constructs of (\ref{eq:funcGram})  we add (demonic) choice $\sqcap$  and patterns:
\[ N ::= M  \mid \LL N,N\RR \mid  N N 
   \mid \lambda x.N  \mid  \plambda x.M\patarrow N \mid  N \sqcap N \] 
In a pattern matching abstraction $\plambda x.M\patarrow N$, pattern $M$ 
is a functional term, 
not just constructors and variables ---recall the rotate example (\ref{eq:rotate}) in Section~\ref{sect:intro}.
The typing rules are 
\[
\frac{\G \tg N : \typ \quad \G \tg N' : \typ}
{\G \tg N\sqcap N' : \typ}
\qquad
\frac{\G,x:\typ ''\tg M:\typ \quad \G,x:\typ ''\tg N:\typ '}  
{\G\tg\plambda x.M\patarrow N:\typ \typarrow \typ '}
\quad
\]
For a term $M$ of functional type, say $\typ\typarrow\typ'$, 
one can think of the term $\plambda x\bdot M x \patarrow x : \typ'\typarrow\typ$
as the converse of $M$.

In this section we write $\se\slot\es$ for the relational interpretation $\R\se\slot\es$. 
Types are interpreted as before, except for using the relational constructs:
$\se \typ \times \typ '\es = \se \typ \es \times \se \typ '\es$ and 
$\se \typ \typarrow \typ '\es = \se \typ \es \homi \se \typ '\es$.
We assume $\Ri\se b\es = \F\se b\es$ for the base types, hence 
if $\typ$ is arrow-free then $\Ri\se\typ\es = \F\se\typ\es$.

For $\G=x_0:\typ _0,\ldots,x_n:\typ _n$, the semantics of a judgement $\G\tg N:\typ $ is
an ideal $\se \G\es \arrow \se \typ \es$ defined in Table~\ref{tab:rsem}.  
\begin{table}[t]
    \caption{The relational semantics $\R\se\slot\es$,
given ideals $\se c:\typ \es :\se\one\es\arrow \R\se\typ\es$.}
    \label{tab:rsem}
\[
\begin{array}{lcl}
\Ri\se \G\tg x:\typ\es & = & \Gr\pi \quad\mbox{where $\pi$ is the appropriate projection} \\
\Ri\se \G\tg c:\typ\es & = & \Gr\pi\co\Ri\se c:\typ\es \mbox{ where $\pi$ is the projection
    $\se\G\es\arrow\se\one\es$} \\ 
\Ri\se \G\tg \LL N,N'\RR:\typ \times \typ ' \es & = & \LL \Ri\se \G\tg N:\typ \es, \Ri\se \G\tg
            N':\typ '\es\RR \\
\Ri\se \G\tg N N':\typ \es & = & \LL \Ri\se \G\tg N:\typ'\typarrow\typ\es, \Ri\se \G\tg
N':\typ'\es\RR \co \api \\ 
\Ri\se \G\tg \lambda x. N:\typ \typarrow \typ '\es & = &  \curi\Ri\se \G,x:\typ \tg N:\typ '\es \\
\Ri\se \G\tg \plambda x.M\patarrow N:\typ \typarrow \typ '\es & = & \\ 
\multicolumn{3}{l}{\qquad
  \LL (\simi^*_{\G} \co \Gr(\F\se\G\tg \lambda x.M\es)\co\rg\co 
  (\simi^*_{\typ}\homi\simi_{\typ''} )) \,,\,
   \Ri\se\G\tg\lambda x.N\es \RR \,\co\, \comp
}
\\[1ex]
\Ri\se \G\tg  N\sqcap N':\typ  \es & = & \Ri\se \G\tg N:\typ \es\cup \Ri\se \G\tg N':\typ \es
    \end{array}
\]
 \end{table}
In addition to demonic choice, $\sqcap$, interpreted by union, one can add an operation for conjunction of specifications, interpreted by intersection which preserves the ideal property.
This is omitted, following \cite{deMoor:Gib}.

As explained later, the semantics of pattern terms uses the functional semantics 
$\F\se\slot\es$ of the pattern.
The table omits semantics of 
$\mathit{fst}$ and $\mathit{snd}$ as the image under $\Gr$ of their functional semantics.
Prop.~\ref{thm:main} gives a condition that is desirable to connect the relational interpretation of a constant with its functional interpretation.
For arrow-free types the condition is simply that $\se c:\sigma \es = \Gr(\F\se c:\sigma\es)$.

To interpret patterns, the idea is to compose the converse of the pattern with the
result term \cite{deMoor:Gib}.
This must be done internally, i.e., using morphisms that model the external operation of converse, and unlike $\Rel$, $\Idl$ lacks converse.  
So, unlike in \cite{deMoor:Gib}, our patterns are restricted to functional terms.  For these, the functional interpretation gives a comap  that can be reversed using the internal opgraph $\rg$ which is analogous to $\gr$.   
This is problematic because $\rg$ is anti-monotonic.  
Before delving into the details of the solution, let us turn aside to consider the application of relational semantics.

\paragraph{Falling short of heap patterns}

Here is a simple realization of heaps.
Let the base types include $\mathit{Ref}$ and $\mathit{Heap}$,
with $\Ri\se \mathit{Ref}\es$ some set, ordered discretely.
Let $\Ri\se \mathit{Heap}\es$ be the set of finite partial maps 
$\Ri\se \mathit{Ref}\es \arrow \Ri\se \mathit{Ref}\es\times \Ri\se \mathit{Ref}\es$, again ordered discretely.
One constant is $cell:\mathit{Ref} \times \mathit{Ref} \times \mathit{Ref} \typarrow \mathit{Heap}$,
interpreted as a total function that forms one-cell heaps, written $\cell{x}{y}{z}$ in Section~\ref{sect:over}.
Another constant is $new:\mathit{Heap}\typarrow \mathit{Ref}$, interpreted as some relation that returns references not in the domain of the heap.
It could be a total function (provided that $\Ri\se \mathit{Ref}\es$ is infinite),
or partial, or not even functional.
The constant $star:\mathit{Heap}\times \mathit{Heap}\typarrow \mathit{Heap}$ is interpreted as a partial function that forms the union, for heaps with disjoint domains, and is otherwise undefined.

These definitions do not yet achieve the hoped-for ability to write patterns as in Section~\ref{sect:over}, because to appear on the left of a pattern term the constants need to have an interpretation in $\Mofun$.   
This might be handled by adding an ``undefined element'' to $\Ri\se \mathit{Heap}\es$, and using a deterministic allocator.
A nicer solution is to move to predicate transformers (Section~\ref{sect:tran}), where the pattern term merely needs an interpretation in $\Idl$, for which the definitions above are fine.

\paragraph{Completing the semantics of pattern terms}

The rest of this section gives the technical details of the semantics of pattern matching.

Recall that the typing rule infers 
$\G\tg\plambda x.M\patarrow N:\typ \typarrow \typ '$ 
from the judgements $\G,x:\typ ''\tg M:\typ $ and $\G,x:\typ ''\tg N:\typ '$.  
The one fine point about typing is that it should enforce, somehow, that there is an interpretation in $\Mofun$ for all constants in $M$. We refrain from formalizing that.  
The semantics is defined using the semantics of the corresponding abstractions, namely 
\[
\begin{array}{l}
\mor{ \F\se \G\tg \lambda x.M:\typ ''\typarrow \typ \es
      }{ \F\se\G\es }{ \F\se \typ ''\es\homm\F\se \typ \es } \quad\mbox{in }\Mofun
\\
 \mor{ \Ri\se \G\tg \lambda x.N:\typ ''\typarrow \typ '\es
      }{ \se\G\es }{ \se \typ ''\es\homi\se \typ '\es } \quad\mbox{in }\Idl
\end{array}
\]
Consider the special case where 
$\G,\typ,\typ',\typ''$ are arrow-free, so the two semantics $\F$ and $\R$ agree on
types. In that case, we want to compose the pattern term with the 
internal opgraph $\mor{\rg}{\se\typ''\es\homm\se\typ\es}{\se\typ\es\homi\se\typ''\es}$
to get 
\begin{equation}
  \label{eq:typ}
 \mor{ \F\se \G\tg \lambda x.M:\typ ''\typarrow \typ \es \co\rg 
\: }{\: \se\G\es }{ \se \typ \es\homi\se \typ ''\es } 
\end{equation}
which, paired with the semantics of $N$, gives 
$ \se\G\es \arrow (\se \typ \es\homi\se \typ ''\es) \times (\se \typ ''\es\homi\se \typ
'\es)  $. 
Following this with the internal composition $\comp$ yields 
what we need:
\begin{equation}
\label{eq:a} 
  \LL\F\se \G\tg \lambda x.M\es \co\rg \,,\, \Ri\se \G\tg \lambda x.N\es \RR \co
  \comp
\::\:\se\G\es \arrow \se \typ \es\homi\se \typ '\es
\end{equation}
This is the gist of the idea.
But $\rg$ is anti-monotonic so $\F\se \G\tg
\lambda x.M\es \co\rg$ need not be an ideal and the other constructs do not apply.  

To solve this problem, we first extend $\gr$ to 
a type-indexed family of  ideals
$\mor{\simi_\typ }{\F\se \typ \es}{\R\se \typ \es}$.
\[
\begin{array}{lll}
\simi_b &=& \Id \quad \mbox{i.e., $\pol_b$, for base types $b$ (and for $\one$)} \\
\simi_{\typ \times \typ '} &=& \simi_\typ  \times \simi_{\typ '} \\
\simi_{\typ \typarrow \typ '} &=& \gr  \co (\simi_\typ ^* \homi \simi_{\typ '} )
\end{array}
\]
For all $\typ $, we have that   $\simi_\typ $ is a comap, because 
$\gr $ is a comap,  the range of $\homi$ is comaps, and
composition preserves comaps.  Thus the map $\simi_\typ ^*$  exists.

Next, we define $\rg$. 
Other elements of the semantics are based solely on structure given by
Propositions~\ref{ax:mo}--\ref{ax:ex}, but for $\rg$ we leave such an axiomatic treatment to
future work. 
For any $B,C$, let $\Rg_{B,C}$ be the restriction of $\Rg$ to the homset
$\Mofun(B,C)$, so that $\Rg_{B,C}$ is a monotonic function $\mor{\Rg_{B,C}}{\Mofun(B,C)}{\postop{\Idl(C,B)}}$.
But the target poset, which can also be written $\postop{C\homi B}$, is ordered
upside-down from what is needed to compose with the result term in a pattern expression.
For any $B$ let $\rev_B$ be the identity function, taken as an anti-monotonic function of type $\op{B}\arrow B$.
Define 
$\mor{\rg}{(B\homm C)}{(C \homi B)}$ as the anti-monotonic function $\Rg_{B,C}\co
\rev_{B\homi C}$.
Thus we have the composite (\ref{eq:typ}) as a relation of the type shown there, though it is not an ideal.
Now we can use the \dt{sandwich lemma}:
For relations  $ A \rTo ^{Q} B \rTo^{R} C \rTo^{S} D $ between posets,
\begin{equation}\label{eq:sand}
\mbox{if } Q,S \mbox{ are ideals then } Q\co R\co S \mbox{ is an ideal }A\arrow D  
\end{equation}
We sandwich $\F\se \G\tg \lambda x.M\es \co\rg$ with ideals based on 
$\simi$ in a way that is needed anyway to reconcile the $\F$ and $\R$ interpretations of 
types. To give the details, we note first the types 
\[
\begin{array}{l}
 \mor{ \R\se \G\tg \lambda x.M':\typ ''\typarrow \typ '\es
      }{ \R\se\G\es }{ \R\se \typ ''\es\homi\R\se \typ '\es } \quad\mbox{in }\Idl
\\
\mor{ \F\se \G\tg \lambda x.M:\typ ''\typarrow \typ \es
      }{ \F\se\G\es }{ \F\se \typ ''\es\homm\F\se \typ \es } \quad\mbox{in }\Mofun
\end{array}
\]
and hence 
$\F\se \G\tg \lambda x.M:\typ ''\typarrow \typ \es \co \rg $ is a relation 
$\F\se\G\es \arrow \F\se \typ \es\homi\F\se \typ'' \es$.
Using appropriate instances of  $\simi$ and $\simi^*$ we obtain 
\[ \simi^*_{\G} \co \Gr(\F\se \G\tg \lambda x.M:\typ ''\typarrow \typ \es)
       \co\rg\co (\simi^*_{\typ }\homi\simi_{\typ''} ) \]
of type
$ \R\se\G\es  \longrightarrow \R\se \typ \es\homi \R\se \typ'' \es$.  This is an ideal, by
the sandwich lemma (\ref{eq:sand}).  
As described earlier, 
for the semantics of
$\plambda x.M\patarrow N$ 
this ideal is paired with the semantics of $N$ and followed
by $\comp$, in accord with (\ref{eq:a}). See Table~\ref{tab:rsem}. Although $\Gr$ could be omitted, 
owing to (\ref{eq:gr}), we use it
because it has a parallel in the predicate transformer semantics (Table~\ref{tab:tsem}).


\section{Lax laws for relational semantics}\label{sect:ilaws}

This section proves a connection between relational and functional semantics,
describes a stronger connection conjectured to hold,
and proves new results that serve as laws of programming.

\paragraph{Connecting the semantics by simulation}

A conservative extension result would show that, 
for a purely  functional term, the relational (or transformer) interpretation is ``the
same'' as the functional one, in some suitable sense.  
Given that the two semantics have different interpretations for arrow types, this needs to involve the embedding between those interpretations, as noted in Section~\ref{sect:intro}.
De Moor and Gibbons \cite{deMoor:Gib} state a strong conservative extension result
for their relational semantics, which we explain later.
For the models in this article, weaker simulation results
suffice to justify program construction by stepwise refinement.

\begin{proposition}\label{thm:main}
\upshape  \cite{Naumann01b}
Suppose that 
$ 
\simi_{\one} \co \R\se c:\typ \es \sps \F\se c:\typ \es \co \simi_\typ  $
for all constants\footnote{If $\G=x_0:\typ _0,\ldots,x_n:\typ _n$, we  write $\simi_{\G}$  for  $\mor{\simi_{(\ldots(\one \times \typ _0)\times \ldots  \typ _n )}}{\F\se\G\es}{\R\se\G\es}$.
}
  $c:\typ $.
Then for all functional terms in context $\G\tg M:\typ$
\[ 
\simi_{\G} \co \R\se\G\tg M:\typ \es \ssps \F\se\G\tg M:\typ \es \co \simi_\typ  
\]
\end{proposition}
The right hand side is the same as 
$\Gr(\F\se\G\tg M:\typ \es) \co \simi_\typ$, owing to (\ref{eq:gr}). 
Owing to the shunting property (\ref{eq:shunt}), there is an equivalent formulation which also 
suggests how to obtain relational interpretations of constants from their functional interpretation:
\[ 
\R\se\G\tg M:\typ \es 
\ssps 
\simi^*_{\G} \co \Gr(\F\se\G\tg M:\typ \es) \co \simi_\typ  
 \] 

Corollary: 
If all types in $\G,\typ $ are arrow-free,
$\R\se\G\tg M:\typ \es \sps \F\se\G\tg M:\typ \es$, 
because $\simi$ is the identity for arrow-free types and $\Gr$ is increasing.
This licenses development by stepwise refinement chains
$\R\se N\es \sps \ldots \sps\R\se M\es \sps \F\se M\es$.  

\paragraph{Connecting the semantics as an equality}

The conservative extension property \cite{deMoor:Gib} involves two additional interpretations of types.
The positive interpretation $\Pos$ replaces every arrow in positive position
by a relation space and every negative arrow by a function space.  The negative interpretation $\Neg$ does the reverse.
For example 
$\Pos\se(b_0\typarrow b_1)\typarrow(b_2\typarrow b_3)\es$ is 
$(\se b_0\es \homm \se b_1\es)\homi(\se b_2\es\homi\se b_3\es)$
for  base types $b_0,\ldots,b_3$.
The definitions are mutually recursive:
\[
\begin{array}{llllll}
\Pos\se \typ\typarrow\typ'\es & = & \Neg\se\typ\es \homi \Pos\se\typ'\es
\qquad & 
\Neg\se \typ\typarrow\typ'\es & = & \Pos\se\typ\es \homm \Neg\se\typ'\es
\\
\Pos\se \typ\times\typ'\es & = & \Pos\se\typ\es \times \Pos\se\typ'\es
& 
\Neg\se \typ\times\typ'\es & = & \Neg\se\typ\es \times \Neg\se\typ'\es
\\
\Pos\se b\es & = & \se b \es
& 
\Neg\se b\es & = & \se b \es
\end{array}
\]
For clarity the connection is stated for the special case that the context has a single variable.
The connection is the equality of the 
the top and bottom paths in this diagram.
\begin{equation}\label{eq:diag:hex}
\begin{diagram}[h=4ex,w=3em] 
          &               & \F\se\typ\es  & \rTo^{\F\se x:\typ \tg M : \typ'\es \quad } & \F\se\typ'\es   &                 &  \\
          & \ruTo^{n2f_\typ} &               &                                &                 & \rdTo^{f2p_{\typ'}} & \\
\Neg(\typ) &               &              &                                &                  &                 & \Pos(\typ') \qquad \\
          & \rdTo_{n2r_\typ} &               &                                &                  & \ruTo_{r2p_{\typ'}} & \\
          &               & \R\se\typ\es  & \rTo^{\R\se x:\typ \tg M : \typ'\es \quad} & \R\se\typ'\es    &                  & \\
\end{diagram}
\end{equation}
To be precise, $\Gr$ should be applied to the upper path.
The upper diagonals are two families of monotonic functions defined by mutual recursion on types.
The lower diagonals are two families of ideals defined similarly.
\[
\begin{array}{llllll}
n2f_{\typ\typarrow\typ'} &=& f2p_\typ \homm n2f_{\typ'} 
& 
f2p_{\typ\typarrow\typ'} &=& (n2f_\typ \homm f2p_{\typ'})\co \Gr 
\\
n2f_{\typ\times\typ'} &=& n2f_\typ \times n2f_{\typ'} 
&
f2p_{\typ\times\typ'} &=& f2p_\typ \times f2p_{\typ'} 
\\
n2f_b &=& \id
& 
f2p_b &=& \id 
\\[2ex]
n2r_{\typ\typarrow\typ'} &=& \gr\co (r2p_\typ \homi n2r_{\typ'}) 
\qquad 
&
r2p_{\typ\typarrow\typ'} &=& n2r_\typ \homi r2p_{\typ'}
\\
n2r_{\typ\times\typ'} &=& n2r_\typ \times n2r_{\typ'} 
&
r2p_{\typ\times\typ'} &=& r2p_\typ \times r2p_{\typ'} 
\\
n2r_b &=& \Id 
&
r2p_b &=& \Id 
\end{array}
\]
Here $\Gr$ is the function $\mor{\Gr}{\Neg\se\typ\es\homm \Pos\se\typ'\es}{\Neg\se\typ\es\homi \Pos\se\typ'\es}$
and $\gr$ is the ideal $\mor{\gr}{\Pos\se\typ\es\homm \Neg\se\typ'\es}{\Pos\se\typ\es\homi \Neg\se\typ'\es}$.

The result conjectured in \cite{deMoor:Gib} is that, for their semantics,
the hexagon equality holds for all functional terms (assuming that it holds for the interpretations of constants).
We conjecture that this also holds in our semantics.\footnote{De Moor and Gibbons 
indicate that they proved the result for beta normal forms.
A possibly relevant observation is that in our $\R$ semantics, application-free functional terms are comaps; applications are interpreted using $\api$ which is not a comap.}

\paragraph{Laws for programming}

Stepwise refinement depends on monotonicity of program constructs.  
Monotonicity of pattern terms is delicate, due to the use of $\rg$, and we leave that issue
to future work on laws for patterns.  
Monotonicity is not addressed in \cite{deMoor:Gib}, perhaps due to difficulties 
mentioned in Section~\ref{sect:ideal}, but it is straightforward to prove the
following for relational terms in $\Idl$.  
\begin{theorem}[monotonicity]
\upshape
Let $C[\slot]$ be any context\footnote{A context $C[\slot]$ is a term with a missing subterm, called the hole, and $C[N]$ is the term with the hole filled by $N$.}
such that the hole does not occur on the left of $\patarrow$ in a pattern. Let $N$ and $N'$ be of suitable type to fill the hole.
Then 
$\R\se \G\tg N:\typ\es\sps  \R\se \G\tg N':\typ\es$ implies 
$\R\se \G'\tg C[N]:\typ'\es\sps  \R\se \G'\tg C[N']:\typ'\es$.
\end{theorem}

Proposition~\ref{thm:main} licenses us to reason about functional terms using functional
semantics.  But the extended language would be of little interest if laws did not carry
over to it.  Laws for products in the relational semantics can be read
directly from corresponding semantic properties, and are left to the reader.  

For lambda expressions the eta law is $\lambda x\bdot N x \sps N$ (for $x$ not free in $N$). 

\begin{theorem}[eta law]
\upshape 
$\R\se\G\tg\lambda x\bdot N x : \typ\typarrow\typ'\es  \sps 
\R\se\G\tg N : \typ\typarrow\typ'\es$ provided that $x$ is not free in $N$.
\end{theorem}
Proof:
\begin{calcu}
  \fln{\R\se\G\tg \lambda x\bdot Nx:\typ\typarrow \typ'\es }
\pln{=}{\curi(\LL \R\se\G,x:\typ\tg N:\typ\typarrow\typ'\es \, ,\,   
                  \R\se\G,x:\typ\tg x:\typ\es\RR \co \api ) }{semantics of $\lambda$ and
                  application} 
\pln{=}{\curi(\LL \pi\co\R\se\G\tg N:\typ\typarrow\typ'\es\,,\, \pi'\RR \co \api) }{
         $x$ not free in $N$, semantics of $x$ ($\pi,\pi'$ are left, right proj.)}
\pln{=}{\curi( (\R\se\G\tg N:\typ\typarrow\typ'\es\times \Id) \co \api )}{def $\times$}
\pln{\sps}{\R\se\G\tg N:\typ\typarrow\typ'\es}{exponent law (\ref{eq:apx})}
\end{calcu}
The second step uses the evident semantics of a typing rule that adds variable
$x$ to the context for a term in which $x$ is not free.  

The semantics validates $(\lambda x.x-x)(0\sqcap 1) = 0$, which  shows
that application is by-value rather than by-name. 
As expected in a by-value calculus, beta conversion is not an equality in 
the relational semantics, e.g.,  $(\lambda x.x-x)(0\sqcap 1) \ss (0\sqcap 1)-(0\sqcap 1) $
is a proper inclusion because the right side includes $0-1$ and $1-0$.
We do get a refinement for this and similar cases. 
The beta law $N[N'/x] \sps (\lambda x\bdot N)N'$ holds provided that $x$ does not
occur in a pattern.  
We say $x$ \dt{occurs in a pattern of} $N$ if it occurs in $M$ for some subterm
$\plambda x\bdot M\patarrow N_1$ of $N$.

\begin{theorem}[beta law]
\upshape \label{thm:betaIdl}
For all relational terms $N$ and $N'$, such that $x$ does not occur in a pattern in $N$, 
$\R\se\G\tg N[N'/x]:\typ\es \: \sps\:  \R\se\G\tg (\lambda x\bdot N)N':\typ\es$.
This is an equality if $N'$ denotes a comap and $N$ is pattern-free. 
\end{theorem}
Note that the equality still allows nondeterminacy in $N$.  

In the rest of this section we omit $\R$.
To prove the theorem, we observe 
\begin{calcu}
\fln{\Ri\se\G\tg (\lambda x\bdot N)N':\typ \es}
\pln{=}{ \LL \cur \Ri\se \G,x: \typ'\tg N:\typ\es\,,\, \Ri\se \G\tg N':\typ'\es\RR \co \ap
       }{semantics of $\lambda$ and application}
\pln{=}{ \LL \Id,\Ri\se \G\tg N':\typ'\es\RR  
       \co (\cur \Ri\se \G,x: \typ'\tg N:\typ\es \times \Id) \co \ap }{product law
       (\ref{eq:prc})} 
\pln{=}{
 \LL \Id, \Ri\se \G\tg N':\typ'\es\RR \,\co\, \Ri\se \G,x: \typ'\tg N:\typ\es 
}{exponent law (\ref{eq:ap}) }
\end{calcu}
The last line  exhibits ``substitution as composition''.
The proof is completed  using a substitution lemma.  While for functional semantics it is
an equality, for ideals it weakens to an inequality.  

\begin{lemma}[substitution]
\label{lem:subst}
\upshape
For all relational terms $N,N_0$ such that $x$ does not occur in a pattern of $N$,
\[ \se\G\tg N[N_0/x]:\typ\es \sps  \LL \Id, \se \G\tg N_0:\typ_0\es\RR\co\se\G,x\ty \typ_0\tg N:\typ\es \]
Equality holds if $N_0$ denotes a comap and $N$ is pattern-free. 
\end{lemma}

An operational interpretation of the inequality goes as follows.
On the right side, $N_0$ is evaluated once, whereas  on the left there may
be multiple occurrences of $N_0$ in $N[N_0/x]$ which give rise to different
nondeterministic choices. 
 
The rest of this section is devoted to the proof of Lemma~\ref{lem:subst}.
The inequality is proved by structural induction on $N$.  
We give the proof in detail, in order to pinpoint exactly what properties are needed
for the inequality and for the equality.  
We elide $\R$ throughout, and also elide some types, including the type of $x$.

\textbf{Case} $N$ is a variable $x$.  Here $N[N_0/x]$ is $N_0$ and we have 
\begin{calcux}
  \flnx{ \LL \Id, \se\G\tg N_0\es\RR \co \se\G,x\tg x\es }
\plnx{=}{ \LL \Id, \se\G\tg N_0\es\RR \co \Gr \pi}{semantics of $x$}
\plnx{(A)\quad =}{ \se\G\tg N_0\es}{product law (\ref{eq:pra}=), $\Id$ comap}
\end{calcux}
Here  (A) marks a step to which we return when proving Lemma~\ref{lem:substt} for
transformer semantics.    

\textbf{Case} $N$ is a choice $N'\sqcap N''$. Straightforward, using that $\co$ distributes
over  $\cup$.  The straightforward cases for constants and pairing are also omitted.

\textbf{Case} $N$ is an application $N'N''$.  
\begin{calcu}
  \fln{\LL \Id, \se\G\tg N_0\es\RR \co \se\G,x\tg N'N''\es }
  \pln{=}{\LL \Id, \se\G\tg N_0\es\RR \co \LL \se\G,x\tg N'\es , \se\G,x\tg N''\es \RR\co \api
        }{semantics of $N'N''$}
  \pln{(B)\quad \ss}{\LL\LL \Id, \se\G\tg N_0\es\RR \co \se\G,x\tg N'\es 
\,,\, 
\LL \Id, \se\G\tg N_0\es\RR \co \se\G,x\tg N''\es \RR\co \api
        }{product law (\ref{eq:prb}), $\co$ monotonic}
\pln{\ss}{\LL \se\G\tg N'[N_0/x] \es,  \se\G\tg N''[N_0/x]\es\RR\co \api
        }{induction, $\co,\LL\slot,\slot\RR$ monotonic}
  \pln{=}{  \se\G\tg (N'N'')[N_0/x] \es }{semantics of application, substitution}
\end{calcu}
Note that step (B) is an equality if $N_0$ denotes a comap, by (\ref{eq:prb}=).  

\textbf{Case} $N$ is an abstraction $\lambda y\bdot N'$.  Clearly the semantics models alpha
conversion, so we can assume without loss of generality that $y$ is distinct from $x$.
And $y$ is not free in $N_0$.
We need to use an isomorphism $p$ that rearranges components $x,y$ of the context.  Such
structural isomorphisms are available thanks to Propositions~\ref{ax:mo} and \ref{ax:emb}.
Precise definitions can be found in semantics texts such as \cite{GunterBook}.
We content ourselves with displaying the
type of $p$ in a diagram for an equation,  that holds for arbitrary
$\mor{R}{\se\G\es}{\se\typ_0\es}$ and any $\typ''$.  
\begin{equation}
  \label{eq:pcom}
\begin{diagram}[w=5em,h=2em]
\se\G\es\times \se\typ''\es &\rTo^{\LL \Id,\pi\RR\quad}&(\se\G\es\times\typ''\es)\times\se\G\es \\
\dTo^{\LL\Id, R\RR\times\Id} & & \dTo_{\Id\times R} \\
(\se\G\es\times \se\typ_0\es)\times\se\typ''\es &\lTo^p & (\se\G\es\times\se\typ''\es)\times\se\typ_0\es 
\end{diagram}  
\end{equation}
Using (\ref{eq:pcom}) we calculate
\begin{calcux}
  \flnx{ \se\G\tg(\lambda y\bdot N')[N_0/x]  \es }
\plnx{=}{ \se\G\tg\lambda y\bdot N'[N_0/x]  \es }{substitution, $x,y$ distinct}
\plnx{=}{ \curi \se\G,y\tg N'[N_0/x] \es }{semantics of $\lambda$}
\plnx{\sps}{ \curi ( \LL \Id, \se \G,y\tg N_0 \es\RR\co\se\G,y,x\tg N'\es) 
}{induction, $\curi$ monotonic}
\plnx{=}{ \curi ( \LL \Id, \pi\co\se \G\tg N_0 \es\RR\co\se\G,y,x\tg N' \es) }{$y$ not free in $N_0$, semantics}
\plnx{=}{ \curi ( \LL \Id, \pi\co\se \G\tg N_0 \es\RR\co p \co \se\G,x,y\tg N' \es) }{$p$
  isomorphism }
\plnx{=}{ \curi ( (\LL \Id, \se \G\tg N_0 \es\RR\times \Id)\co  \se\G,x,y\tg N' \es) 
}{(\ref{eq:pcom}) with $R:=\se\G\tg N_0\es$}
\plnx{(C)\quad \sps}{ \LL \Id, \se \G\tg N_0 \es \RR \co  \curi\se\G,x,y\tg N' \es)
 }{exponent law (\ref{eq:curnat}) }
\plnx{=}{ \LL \Id, \se \G\tg N_0 \es \RR \co  \se\G,x\tg \lambda y\bdot N' \es) }{semantics of $\lambda$}
\end{calcux}
Step (C) is an equality if $N_0$ denotes a comap, by (\ref{eq:curnat}=).  

\textbf{Case} $N$ is a pattern term 
$\plambda y\bdot M\patarrow N'$. 
Assume w.l.o.g.\ that $x$ is distinct from $y$.
Here $N_0$ must be a functional term in order for the
substitution to yield a typable pattern term.
We write $\sims$ to abbreviate $\simi$, and a huge comma $\bigcomma$ to aid parsing
complicated pairings. We also elide $\G$, and the type of $x$, on the left of $\tg$ throughout.

\begin{calcu}
  \fln{ \LL \Id, \se\tg N_0\es\RR \co \se x\tg \plambda y\bdot M\patarrow N'\es }
\pln{=}{ \LL \Id, \se\tg N_0\es\RR 
\co \LL \sims^*\co\F\se x\tg \lambda y\bdot M\es\co\rg\co(\sims^*\homi\sims)
\bigcomma
 \se x\tg\lambda y\bdot N'\es\RR \co\comp 
  }{semantics of $\plambda$}
\pln{(D)\, \ss}{ 
\Bigl\LL
\LL \Id, \se\tg N_0\es\RR 
\co \sims^*\co\F\se x\tg \lambda y\bdot M\es\co\rg\co(\sims^*\homi\sims)
\bigcomma
\LL \Id, \se\tg N_0\es\RR \co \se x\tg\lambda y\bdot N'\es
\Bigr\RR 
\co\comp
  }{by (\ref{eq:prb}) }
\pln{\ss}{ 
\Bigl \langle
\LL \Id, \se\tg N_0\es\RR 
\co \sims^*\co\F\se x\tg \lambda y\bdot M\es\co\rg\co (\sims^*\homi\sims)
\bigcomma
\se \tg\lambda y\bdot N'[N_0/x] \es
\Bigr\rangle
 \co\comp   }{induction}
\pln{\ss}{ \LL\sims^*\co \F\se \tg \lambda y\bdot M[N_0/x]\es\co\rg\co(\sims^*\homi\sims)
\bigcomma
\se \tg\lambda y\bdot N'[N_0/x] \es\RR \co\comp
  }{claim (\ref{eq:claim}) below, and monotonicity of $(\co)$ and $\LL\slot,\slot\RR$}
\pln{=}{ \se \tg (\plambda y\bdot M\patarrow N')[N_0/x]\es }{substitution, semantics of $\plambda$}
\end{calcu}
Step (D) is an equality if $N_0$ denotes a comap, by (\ref{eq:prb}=). The claim is
\begin{equation}
  \label{eq:claim}
 \sims^*\co \F\se \tg \lambda y\bdot M[N_0/x]\es 
\sps \LL \Id, \se\tg N_0\es\RR \co \sims^*\co \F\se x\tg \lambda y\bdot M\es   
\end{equation}
If $x$ is not free in pattern $M$ then $\lambda y\bdot M[N_0/x]$ is $\lambda y\bdot M$ and the
claim can be proved using properties of the left projection $\pi$ as follows.
\begin{calcux}
  \flnx{ \LL \Id, \se\tg N_0\es\RR \co \sims^*\co \F\se x\tg \lambda y\bdot M\es }
\plnx{=}{\LL \Id, \se\tg N_0\es\RR \co \sims^*\co \pi\co\F\se \tg \lambda y\bdot M\es }{
$x$ not free in $\lambda y\bdot M$, semantics}
\plnx{=}{\LL \Id, \se\tg N_0\es\RR \co (\sims^*\times \sims^*)\co \pi\co\F\se \tg \lambda
  y\bdot M\es 
}{
def $\sims$, $\times$ preserves comaps}
\plnx{(E)\quad\ss}{\LL \Id, \se\tg N_0\es\RR \co \pi\co \sims^*\co \F\se \tg \lambda
  y\bdot M\es }{$\pi$ lax natural, from (\ref{eq:pra})}
\plnx{(F)\quad\ss}{ \sims^*\co \pi\co\F\se \tg \lambda y\bdot M\es }{(\ref{eq:pra}) and unit
  law}  
\end{calcux}
Because $\times$ preserves comaps, $\sims_{\G,x}^* = \sims_\G^* \times \sims_x^*$.  
Step (F) is an equality if $N_0$ denotes a comap, but step (E) is not.  

The lemma holds as an equality if $N_0$ is a comap and $N$ has no patterns.  
(Under those conditions, the inductive steps become equalities.)

\medskip

If $x$ occurs in $M$, a natural attempt goes as follows.
By the substitution lemma for $\F$, which is an equality, (\ref{eq:claim}) is equivalent to 
\[ \sims^*\co \LL\Id , \F\se \tg N_0\es \RR\co \F\se x\tg \lambda y\bdot M\es 
\sps \LL \Id, \se\tg N_0\es\RR \co \sims^*\co \F\se x\tg \lambda y\bdot M\es \]
This follows by monotonicity of $(\co)$ from 
$ \sims^*\co \LL\Id , \F\se \tg N_0\es \RR \sps \LL \Id, \se\tg N_0\es\RR \co \sims^*$.
Here we are in trouble. Roughly speaking, this asks for the functional semantics of 
$N_0$ to  contain the relational semantics, which is the reverse of  
Proposition~\ref{thm:main}.

\section{Transformers and transformer semantics}\label{sect:tran}

Predicate transformers are often taken to be monotonic functions on powersets, but here we
use updeal lattices.  
The reason is that if powersets are used instead of updeal lattices, the
internal hom does not carry a well behaved exponent structure.  
For example, we need the associated functor to preserve identities, which fails with powersets.
This section begins with a brief explanation of why updeal lattices are sensible in
programming terms. Then it proceeds to describe the model and give the semantics.
Results not proved here can be found in \cite{Naumann98} or \cite{Naumann01b}.

Ordered data types are needed even in first order languages, 
if  extensible record or object types are admitted.
To see the significance of ordering for imperative programs, consider an expression
refinement  $e\pol e'$. 
This does not imply  the command refinement $x:=e \rt x:=e'$,
because that requires $wp(x:=e)\phi\ss wp(x:=e')\phi$ for all postconditions $\phi$, which
fails for postcondition $x=e$.
A more reasonable postcondition is $e\pol x$, which  is upward
closed in $x$.  To give a specification that exactly characterizes the assignment $x:=e$,
we can use auxiliary $y$ in precondition
$y\pol e$ and postcondition $y\pol x$.  These predicates are closed upward in the state
variable $x$ and downward in $y$.  In general, $pre,post$ in the span (\ref{eq:span_s})
can be taken to be ideals from auxiliary state to program state.  


Weakest-precondition functions map predicates on final states to
predicates on initial states, so notation and terminology is most perspicuous if we use
an opposite category.  We also follow convention in predicate transformer semantics and
order predicates by $\ss$; we write $\Uop A$ for the lattice of updeals on $A$, ordered by
$\ss$, as opposed to $\U A$ ordered by $\sps$.  
We define  $\Tran$ to have all posets as objects; and the homset 
$\Tran(A,B)$ is just $\Mofun(\Uop B,\Uop A)$, i.e., $\Mofun^{op}(\Uop A,\Uop B)$.  
Following the convention in refinement calculi, the symbol $\rt$ is used for the
ordering, so $t\rt t'$ iff $t\beta \ss t'\beta$ for all $\beta\in \Uop B$.  
  
Composition in $\Tran$ is
just functional composition, for which we write $\Sc$ so that 
for $t$ in $\Tran(A,B)$ and $u$ in $\Tran(B,C)$ we have $t \Sc u = u\co t$.  
The identity on $A$ is the identity function $id_{\Uop A}$.  
For any ideal $\mor{R}{A}{B}$, the \dt{universal image}
$\A R\in \Tran(A,B)$ is 
defined by $\A R = \Uop(\Ni/R)$, using relational quotient.
At the level of points, 
$a\in \A R X$ iff $\all b \bdot aRb \Rightarrow b\in X$. 
The universal image of an ideal is universally conjunctive.
Moreover, $\A$ is 
refinement injective: $R\sps S \equiv \A R \rt \A S$. Thus it preserves demonic choice:
$\A(R\cup S) = \A R \sqcap \A S$ where $\sqcap$ is pointwise intersection.  Later we use
$\sqcup$ for pointwise union, which models angelic choice.
Define  $\E R$  to be the direct image of $R$, but as a function $\Uop A\arrow\Uop B$ 
and so distinguished from $\mor{\U R}{\U A}{\U B}$.  
Note that $\E$ is a monotonic functor $\Idl^{co\,op}\arrow\Tran$.
A crucial fact is that for any
$R$, $\A R$ is a map with comap $\E R$.  
Thus $\A$ is onto maps. This is an unfortunate clash of terminology as $\Gr$ plays the
same role as $\A$ but $\Gr$ embeds onto comaps.

A \dt{bimap} is a map that is also a comap.  
For any monotonic function $f$,  $\A(\Gr f)$ is a bimap in $\Tran$.  
Sometimes we  omit $\Gr$ and write simply $\A f$.
The situation looks as follows, using fishtail arrows for embeddings and fishhooks
for inclusions.
\[ 
\begin{diagram}[w=5em,h=2em]
\Mofun & \rEmbed_{\Gr} & \mathbf{Comaps}(\Idl) & \rInto & \Idl      &  & \\
       &               & \dEmbed^{\A}             &        & \dEmbed_{\A} & & \\
       &               & \mathbf{Bimaps}(\Tran)& \rInto & \mathbf{Maps}(\Tran) & \rInto &
       \Tran \\
\end{diagram}  
\]
In fact the vertical arrows are categorical equivalences.

A transformer $\mor{t}{A}{B}$ 
is \dt{strict} if $t\emptyset = \emptyset$ (i.e., $t(\mathit{false})=\mathit{false}$) 
and \dt{costrict} if $t B = A$ (i.e., $t(true)=true$, which expresses program termination).
Comaps are strict and maps are costrict.

Cartesian product of underlying posets gives a weak product of predicate transformers; we
overload $\times$ and $\LL,\RR$ but write $\A\pi$ explicitly for the projection lifted
from $\Mofun$.  
\begin{proposition}\label{ax:tprod}
\upshape 
$\times$ is monotonic and for all $t,u,v,w$, 
  \begin{eqnarray}
\label{eq:tproj}
\LL t,u\RR \Sc \A \pi &\rt& t \quad\mbox{if $t,u$ strict} \\
\label{eq:tprojc}
\LL t,u\RR \Sc \A \pi &\rf& t \quad\mbox{if $t,u$ costrict} \\
\label{eq:peta}
 \LL t\Sc\A\pi\,,\, t\Sc\A\pi'\RR &\rt& t \quad\mbox{if $t$ map}  \\
(t\Sc u)\times (v\Sc w) &\rt & (t\times v)\Sc(u\times w) 
\quad(=, \mbox{ if $u,w$ maps}) \\
\label{eq:prx}
\LL t\Sc u \,,\, v\Sc w\RR &\rt& \LL t,v\RR\Sc (u\times w) 
\quad(=, \mbox{ if $u,w$ maps}) \\
\label{eq:txx}
\LL t\Sc u\,,\, t\Sc v\RR &\rt& t\Sc \LL u,v\RR  \quad \mbox{if $t$ map} 
\quad(=, \mbox{ if $t$ bimap}) 
  \end{eqnarray}
\end{proposition}

Adequacy of predicate transformers as a model of nondeterminacy and divergence together
is reflected in the behavior of products.  As in $\Idl$, the projection law
(\ref{eq:tproj}) is not an equality in general (intuitively, $u$ could diverge).
Even the inequality depends on absence of miracles.  The reverse inequality holds for
costrict transformers and \emph{a fortiori} for maps.
The other defining law for products of functions also weakens
to (\ref{eq:peta}), reflecting nondeterminacy as with relations.
The side condition that $t$ is a map expresses the absence of angelic nondeterminacy;
in fact the condition can be weakened to positive conjunctivity.
The inequality (\ref{eq:txx}) holds so long as $t$ is finitely conjunctive, and the
reverse $\rf$ holds if  $t$ a comap.

As with products, the exponent structure is very lax.
The exponent object $B\homt C$ is $\Tran(B,C)$, ordered by $\rt$.
Definitions for currying and application can be found in \cite{Naumann98} and \cite{Naumann01b}; 
we only need the following.
\begin{proposition}\label{ax:specex} 
\upshape \cite{Naumann98}
$\mor{\homt }{\Tran^{op}\times\Tran}{\Tran}$ is a monotonic functor and 
for all $t,u,v$ 
\begin{eqnarray}
\curt\, t \mbox{ is a bimap} & &\\
\label{eq:ap_s}
(\curt~ t \times \Id )\Sc \apt &=& t\\  
\label{eq:apx_s}
\curt((t\times \Id)\Sc \apt) &\rt&  t \quad\mbox{if $t$  map}
\quad (=, \mbox{if $t$  bimap})\\
\label{eq:curnat_s}
\curt((t\times \Id )\Sc u) & \rt & t\Sc \curt~ u \quad\mbox{if $t$ map}
\quad(=, \mbox{if $t$ bimap})  \\
\curi((\Id\times t)\Sc u\Sc v)&=&\curt~ u\Sc (t\homt v)
\end{eqnarray}
\end{proposition}

The analog of the internal graph functor $\gr$ is the internal universal image
$\mor{\univ}{B\homi C}{B\homt C}$
defined by
$ \univ = \curt(\A \api) $.  
It is a bimap, being in the image of $\curt$.
Just as $\gr$ is used in 
(\ref{eq:emac}), $\univ$ expresses how 
$\curt$ and $\apt$ extend their relational counterparts: 
\[
\curt (\A R) = \A(\curi R) \Sc \univ  \qquad\quad
(\univ \times\Id)\Sc \apt = \A\,\api 
\quad .
\]

\paragraph{Transformer semantics}
Types are interpreted as before, except for using the  constructs of $\Tran$: 
$\se \typ \times \typ '\es = \se \typ \es \times \se \typ '\es$ and 
$\se \typ \typarrow \typ '\es = \se \typ \es \homt \se \typ '\es$.
We assume $\se B \es = \R\se B\es = \F\se B\es$ for the base types.
We extend $\univ $ to a simulation  $\mor{\psim_\typ }{\R\se \typ \es}{\PT\se \typ \es}$ 
by defining these morphisms in $\Tran$:
\[
\begin{array}{lll}
\psim_B &=& \Id \mbox{ for base types $B$ (and for $\one$)} \\
\psim_{\typ \times \typ '} &=& \psim_\typ  \times \psim_{\typ '} \\
\psim_{\typ \typarrow \typ '} &=& \univ  \Sc (\psim_\typ ^* \homt \psim_{\typ '} )
\end{array}
\]
Because  $\univ$ is a bimap and the range of $\curt$ is a bimap, each $\psim$ is a bimap,
and $\psim^*$ denotes the corresponding map.

We augment the language of terms with angelic choice $\sqcup$, using a new syntactic
category $P$.  Patterns are now relational terms:
\[ P ::= N  \mid \LL P,P\RR \mid P P
   \mid \lambda x.P \mid  \plambda x.N\patarrow P \mid  P \sqcap P\mid P\sqcup P\] 
The semantics is in Table~\ref{tab:tsem}.
\begin{table}[t]
    \caption{The predicate transformer semantics $\PT\se\slot\es$,
given transformers $\se c:\typ \es: \one \arrow \PT\se\typ\es$.
}
    \label{tab:tsem}
\[    \begin{array}{lcl}
\PTi\se \G\tg x\es & = & \A\pi \quad\mbox{where $\pi$ is the appropriate projection} \\
\PTi\se \G\tg c\es & = & \A\pi\Sc\PTi\se c:\typ\es \mbox{ where $\pi$ is the projection
    $\se\G\es\arrow\se\one\es$} \\ 
\PTi\se \G\tg \LL P,P'\RR:\typ \times \typ ' \es & = & \LL \PTi\se \G\tg P:\typ \es, \PTi\se \G\tg
            P':\typ '\es\RR \\
\PTi\se \G\tg P P':\typ \es & = & \LL \PTi\se \G\tg P:\typ'\typarrow\typ\es, \PTi\se \G\tg
P':\typ'\es\RR \Sc \apt \\ 
\PTi\se \G\tg \lambda x. P:\typ \typarrow \typ '\es & = &  \curt\PTi\se \G,x:\typ \tg P:\typ '\es \\
\PTi\se \G\tg \plambda x.N\patarrow P:\typ \typarrow \typ '\es & = & \\ 
\multicolumn{3}{l}{\qquad
  \LL \psim^*_{\G} \Sc \A(\R\se\G\tg \lambda x.N\es) \Sc \exi
   \Sc (\psim^*_{\typ ''}\homt\psim_\typ )  \:,\: 
   \PTi\se\G\tg\lambda x.P\es \RR \,\Sc\, \comp
}
\\[1ex] 
\PTi\se \G\tg  P\sqcap P':\typ  \es & = & \PTi\se \G\tg P:\typ \es\sqcap \PTi\se \G\tg P':\typ \es\\
\PTi\se \G\tg  P\sqcup P':\typ  \es & = & \PTi\se \G\tg P:\typ \es\sqcup \PTi\se \G\tg P':\typ \es
    \end{array}
\]
 \end{table}
For semantics of patterns we overload the name $\comp$ for the internal composition in
$\Tran$.  We also use the internal existential image  $\exi$, which is
analogous to $\rg$ for the relational semantics.  The analogy includes the annoying fact
that it is an order-reversing function.   For any $B,C$ let $\E_{B,C}$ be the restriction
of $\E$ to the homset $\Idl(B,C)$.  Thus it is a monotonic function 
$\mor{\E_{B,C}}{B\homi C}{\postop{C\homt B}}$, which yields 
anti-monotonic function $\mor{(\E_{B,C}\co \rev)}{B\homi C}{C\homt B}$.
Here $\rev$ is the order-reversing identity function
$\mor{\rev_{C\homt B}}{\postop{C\homt B}}{C\homt B}$
mentioned a few lines before (\ref{eq:sand}).

We defined $\A$ only for ideals, but in fact for any $R$ with $\rpol\co R\ss R$, the
inverse image sends updeals to updeals.  
Taking $R$ to be $\E_{B,C}\co\rev$ and using the order $\sps$ on $B\homi C$ we define
$\exi$ to be $\A(\rsps\co\E_{B,C}\co\rev)$ which is a morphism in 
$\Tran(B\homi C, C\homt B)$.  The rest of the semantics for patterns parallels the
relational semantics.
Any constant that occurs on the left of a pattern needs to have an interpretation in $\Idl$, just as the relational semantics requires a functional interpretation of such constants.


\section{Lax laws for transformer semantics}\label{sect:tlaws}

The basic connection between transformer semantics and ideals is similar to 
Prop.~\ref{thm:main}.

\begin{theorem}[\cite{Naumann01b}] 
\upshape
Suppose that 
$ 
\psim_{\one} \Sc \PT\se c:\typ \es \rt \A(\R\se c:\typ \es) \Sc \psim_\typ  $
for all constants $c:\typ $.
Then for all relational terms $N$
\[ 
\psim \Sc \PT\se \G\tg N:\typ\es \rt \A(\R\se\G\tg N:\typ\es )\Sc \psim
\]
\end{theorem}
If all types in $\G,\typ $ are arrow-free then 
$\PT\se\G\tg N:\typ \es \rt \A(\R\se\G\tg N:\typ \es)$.
The result licenses development by stepwise refinement, that is,  in a chain
$\PT\se P\es \rt \ldots \rt\PT\se N\es \rt \A(\R\se N\es)$ ending with a ``program'' that has
only demonic nondeterminacy. 


\paragraph{Refinement laws}

As in the relational semantics, it is straightforward to prove the following monotonicity
result for transformer terms.   
\begin{theorem}[monotonicity] 
\upshape
For any context $C[\slot]$ in which terms $P$ and $P'$ may occur, except to the left of
$\patarrow$ in pattern terms, 
$\PT\se \G\tg P:\typ\es\sps  \PT\se \G\tg P':\typ\es$ implies 
$\PT\se \G'\tg C[P]:\typ'\es\sps  \PT\se \G'\tg C[P']:\typ'\es$.
\end{theorem}

Laws for products in  predicate transformer semantics can be read
directly from corresponding semantic properties, e.g.,  (\ref{eq:tproj}) and
(\ref{eq:peta}) are beta and eta laws for products in $\Tran$. 
For exponents, the eta law is a conditional refinement:
$\lambda x\bdot P x\rt P$ if $P$ is a map (and $x$ not free in $P$).

\begin{theorem}[eta law]
\upshape
$\PT\se\G\tg\lambda x\bdot P x : \typ\typarrow\typ'\es \rt 
\PT\se\G\tg P : \typ\typarrow\typ'\es$ if $P$ is a map and $x$ is not free in $P$.
\end{theorem}
Proof: We have 
$ \PT\se \G\tg \lambda x\bdot P x \es 
= \curt((\PT\se \G\tg P:\typ\es \times \Id)\Sc \apt) $
by semantics.
If $P$  denotes a map, we can apply law (\ref{eq:apx_s}) to obtain 
$\lambda x\bdot P x\rt P$.  

\begin{theorem}[beta law]
\label{thm:betat}
\upshape
For all $\G\tg P$ and all $\G\tg P'$ such that $\PT\se\G\tg P'\es$ is a map,  
\[ \PT\se\G\tg P[P'/x]:\typ\es \: \rt\:  \PT\se\G\tg (\lambda x\bdot P)P':\typ\es \]
provided $x$ does not occur in a pattern of $P$.
This is an equality if $P'$ denotes a bimap and $P$ is pattern-free.
\end{theorem}
In particular, $P$ may have angelic and demonic nondeterminacy.
Note that $(\slot\Sc t)$ distributes over both  $\sqcap$ and $\sqcup$,
for any $t$.  Also $\E(R\cup S) = \E R\sqcup \E S$.  As a consequence, we have 
$ \plambda x \bdot N\sqcap N' \patarrow P =
(\plambda x \bdot N \patarrow P) \sqcup (\plambda x \bdot N' \patarrow P )$
for all $N,N',P$.  This shows the angelic nature of patterns.

The proof of Theorem~\ref{thm:betat} is like the proof of Theorem~\ref{thm:betaIdl}.  We show
\[ \PT\se \G\tg(\lambda x\bdot P)P':\typ\es 
= \LL \Id, \PT\se \G\tg P':\typ'\es\RR \,\Sc\, \PT\se \G,x: \typ'\tg P:\typ\es \]
using (\ref{eq:prx}=) and (\ref{eq:ap_s}) instead of the corresponding laws for $\Idl$. 
The proof is completed by appeal to the following.
\begin{lemma}[substitution]
\label{lem:substt}
\upshape
For all $P,P_0$ such that $\PT\se\G\tg P_0\es$ is a map
and $x$ does not occur in a pattern in $P$,
\[  \PT\se\G\tg P[P_0/x]:\typ\es \rt 
  \LL \Id, \PT\se \G\tg P_0:\typ_0\es\RR\Sc\PT\se\G,x\ty \typ_0\tg P:\typ\es \]
This is an equality if $P_0$ denotes a bimap and $P$ is pattern-free.
\end{lemma}

The proof is similar to that for Lemma~\ref{lem:subst}, using corresponding properties of
product and exponent in $\Tran$. We do not repeat the calculations; the only
differences are steps marked with letters and we discuss these in turn below. 
We refer to the identifiers $N,N'$ in those calculations, which correspond to $P,P_0$ in
the current proof.
In each case, the step goes through thanks to $P_0$ denoting a map.

For the case that $N$ is $x$, the step
(A) is an inequality $\rf$ thanks to (\ref{eq:tprojc}),
provided  $\PT\se\G\tg N'\es$ is a map. 
For application, the step marked (B) is an inequation $\rf$ if $N'$ denotes a map,
thanks to (\ref{eq:txx}).
In fact costrictness suffices for (A) and positive conjunctivity for (B); together
these are the properties of a map, i.e. universal conjunctivity.
For application, the step marked (C) is an inequality $\rt$ if $N'$ denotes a map, by
(\ref{eq:curnat_s}) (and it is $\rt$ that is needed here, because the calculation is in
the reverse direction from the prior ones).  

For $N$ a pattern term, the step marked (D) goes through by (\ref{eq:txx})
if $N'$ denotes a map, and steps (E) and (F) can be taken using (\ref{eq:tprojc}).  

The proof for demonic choice is straightforward, using that  $(t\Sc \slot)$ distributes
over arbitrary $\sqcap$ if $t$ is a map, and the $t$ in this case is $\PT\se\G\tg
N'\es$.  The proof for angelic choice uses the  fact that $(t\Sc u)\sqcup(t\Sc v) \rt 
t\Sc(u\sqcup v) $ for all $t,u,v$ (by monotonicity of $t$).  

If $N'$ denotes a bimap, the bimap conditions are met for steps (A), (B), and (C),
and bimaps distribute over both $\sqcup$ and $\sqcap$ on the left.  The lemma holds as an
equality for pattern-free $N$ and bimap $N'$.

\section{Discussion}\label{sect:disc}

Following de Moor and Gibbons \cite{deMoor:Gib}, we have shown how to internalize the 
span factorizations of ideal relations and of predicate transformers, and to use these models for two semantics of lambda terms extended with non-injective and non-total pattern matching.
We have shown that not only the language of lambda calculus  but also its basic laws  are
available, although in weakened form.
One notable result is that beta refinement
$P[P'/x]\rt (\lambda x\bdot P)P'$ holds in the transformer model even if $P$ combines 
demonic and angelic nondeterminacy and $P'$ has demonic nondeterminacy.

The language of de Moor and Gibbons includes fixpoints. Our models support recursive definitions, because homsets are complete lattices, but we leave thorough investigation as future work.  
The authors say ``the semantics we have sketched leaves many questions unanswered''.
It would be particularly interesting to check which operations are monotonic with respect to their refinement order, and whether terms in patterns need to be restricted as in our models.

Here are some other open problems.
\begin{itemize}
\item What are the interesting general laws for patterns?
One way to investigate would be to focus on heap patterns: design a subset of those that is easy to compile, and find requisite simplification laws. 

\item To investigate pattern matching for heap operations, 
one could try to derive a version of $repmin$ with shared objects.
Another example close at hand would be to derive conventional code for \emph{in situ} list reversal, starting from the abstract version sketched in Section~\ref{sect:over}.
A further exercise would be to specify and derive a program that maps an operation over a 
list of root pointers to disjoint heap structures.  For example, map $repmin$ over a list of trees.
The pattern could use a confined separation operator along the lines of Wang et al~\cite{WangBO08}.

\item For a usable calculus of imperative programs it seems desirable to avoid explicit threading of state through expressions. 
This suggests combining pattern terms with a monad to encapsulate state, perhaps drawing on ideas from Hoare Type Theory~\cite{NanevskiMB08}.

\item Our presentation emphasizes algebraic structure: the proofs and constructions are pointfree and based on Propositions~\ref{ax:mo}--\ref{ax:ex},\ref{ax:tprod},\ref{ax:specex} ---with a few exceptions.  The semantic definitions for pattern terms rely on manipulation of non-monotonic functions $\rg$ and $\exi$.
The proof of (\ref{eq:curi-uncuri}) is not pointfree.
The problem is to fix these blemishes so that the Propositions can be taken as axioms.

\item The last problem is to prove the hexagon equality (\ref{eq:diag:hex}),
an attractive chiasmus connecting $\F$ with $\R$.  And do the same for $\R$ and $\PT$.
For the semantics of~\cite{deMoor:Gib},
Benton sketched a proof for all functional terms in an unpublished note (2001) \cite{NickBentonPriv}, using a form of logical relation.  
In 2015, according to personal communication with the Benton, de Moor, and Gibbons, there has been no further development, but also little doubt that the result should hold.  
\end{itemize}

\paragraph{Acknowledgments}

Thanks to Oege de Moor and Jeremy Gibbons for their stimulating paper and 
enlightening discussions.  
Many thanks to three anonymous reviewers who provided extensive detailed suggestions that greatly improved the presentation and sharpened my understanding on various points.
Thanks to 
Mounir Assaf,
Anindya Banerjee, 
Nick Benton, 
Steve Bloom, 
Andrey Chudnov,  
Joe Morris,
and 
Jos\'{e} Nuno Oliveira for helpful feedback at various stages of the work reported here.

Special thanks to Jos\'{e} for eloquently advocating and teaching pointfree relation algebra 
and for celebrating my student's remark that those who like maths also enjoy music and maps.
Following the symbols as they danced along the backward arrows of predicate transformers, 
I arrived at a canon cancrizans of co/maps which I offer in celebration 
of the work of Jos\'{e} Nuno Oliveira.

The author acknowledges partial support from NSF award CNS-1228930.

\bibliographystyle{alpha}


\newcommand{\etalchar}[1]{$^{#1}$}

\end{document}